\documentclass[fleqn,usenatbib]{mnras}

\usepackage{newtxtext,newtxmath}

\usepackage[T1]{fontenc}

\DeclareRobustCommand{\VAN}[3]{#2}
\let\VANthebibliography\thebibliography
\def\thebibliography{\DeclareRobustCommand{\VAN}[3]{##3}\VANthebibliography}

\usepackage{graphicx}	
\usepackage{amsmath}	
\usepackage{multirow}
\usepackage{orcidlink}
\definecolor{mod}{rgb}{0,0.0,0}

\newcommand{\ud}{\mathrm{d}}
\newcommand{\ug}{\mathrm{g}}

\newcommand{\pdbyd}[2]{\frac{\partial#1}{\partial#2}}

\graphicspath{{./Figures/}}

\title[Thermal Feedback from Pebble Accretion]{The Role of Thermal Feedback in the Growth of Planetary Cores by Pebble Accretion in Dust Traps}

\author[D. P. Cummins \& J. E. Owen]{
Daniel P. Cummins$^{\orcidlink{0000-0003-2177-6521}}$\thanks{E-mail: daniel.cummins17@imperial.ac.uk (DPC)}
and James E. Owen$^{\orcidlink{0000-0002-4856-7837}}$
\\
Imperial Astrophysics, Imperial College London, Blackett Laboratory, Prince Consort Rd, London, SW7 2AZ, UK
}

\date{Accepted XXX. Received YYY; in original form ZZZ}

\pubyear{2024}

\begin{document}
\label{firstpage}
\pagerange{\pageref{firstpage}--\pageref{lastpage}}
\maketitle

\begin{abstract}
High-resolution millimetre-imaging of protoplanetary discs has revealed many containing rings and gaps. These rings can contain large quantities of dust, often in excess of $10 M_\oplus$, providing prime sites for efficient and rapid planet formation. Rapid planet formation will produce high accretion luminosities, heating the surrounding disc. We investigate the importance of a planetary embryo's accretion luminosity by simulating the dynamics of the gas and dust in a dust ring, accounting for the energy liberated as a resident planetary embryo accretes. The resulting heating alters the flow structure near the planet, increasing the accretion rate of large, millimetre-to-centimetre-sized dust grains. We show how this process varies with the mass of dust in the ring and the local background gas temperature, demonstrating that the thermal feedback always acts to increase the planet's mass. This increase in planet mass is driven primarily by the formation of vortices, created by a baroclinic instability once the accreting planet heats the disc significantly outside its Hill radius. The vortices can then migrate with respect to the planet, resulting in a complex interplay between planetary growth, gap-opening, dust trapping and vortex dynamics. Planets formed within dust traps can have masses that exceed the classical pebble isolation mass, potentially providing massive seeds for the future formation of giant planets. Once pebble accretion ceases, the local dust size distribution is depleted in large grains, and much of the remaining dust mass is trapped in the system's L$_5$ Lagrange point, providing potentially observable signatures of this evolution. 
\end{abstract}




\section{Introduction}
Many protoplanetary discs have been observed to contain narrow, dust-rich rings \citep[e.g.][]{hltau15,andrews16,dsharp1,long18,VanderMarel2019}. There is compelling evidence to suggest that these are gas-pressure enabled dust traps \citep[e.g.][]{dsharp6,Rosotti2023,Pizzati2023,Lee2024}, which inhibit large dust grains from drifting towards the inner disc \citep[e.g.][]{pinilla12a,rosotti20,Kalyaan2021}. The origin of these dust traps is still uncertain: they may arise from a multitude of processes \citep[e.g.][]{VanderMarel2018,Andrews2020} such as planet-carved gaps \citep[e.g.][]{rice06,zhu12,Owen2014,Bae2017,Zhang2018,Booth2020}, photoevaporation \citep[e.g.][]{alexander_armitage07,owen19}, snow-lines \citep[e.g.][]{kretke_lin07,okuzumi16,Hu2019,owen20} or (non-ideal) MHD effects \citep[e.g.][]{johansen09,Flock2015,Hu2022,Hsu2024}. This concentration of large dust grains can promote processes thought to be responsible for the formation of planetary seeds, such as the streaming instability \citep[e.g.][]{youdin_goodman05,carrera21}, although there are challenges \citep[e.g.][]{Carrera2022}. Furthermore, the availability of large dust grains in the vicinity of a planetary seed makes such a dust trap the ideal location for growth into a planetary core \citep[e.g.][]{liu19,morbidelli20,cummins22,Jiang2023,Sandor2024,Pierens2024}. It is, therefore, felicitous to consider the possibility of planet formation within dust rings.

\begin{figure}
    \centering
    \includegraphics[width=\columnwidth, trim={0, 13.cm, 36.6cm, 0},clip]{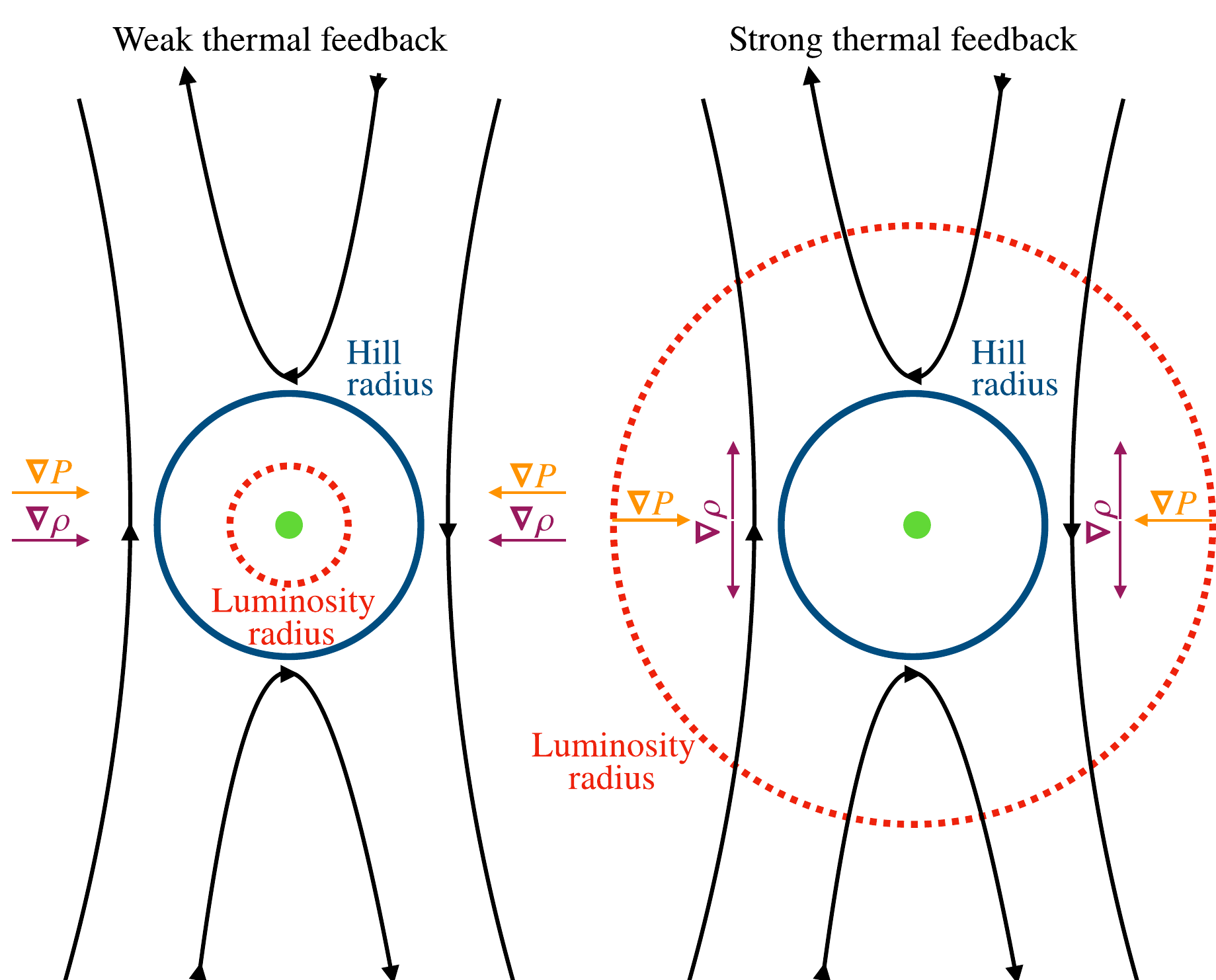}
    \caption{Schematic view in a frame co-rotating with a planet at a pressure maximum, demonstrating how strong thermal feedback from pebble accretion can generate vortices, based on the idea of \citet{owen17}. The black solid lines show gas velocity streamlines around the planet (indicated by a solid green circle). Vorticity is sourced by regions where $\boldsymbol\nabla P \times \boldsymbol\nabla \rho \ne 0$. In the case of weak thermal feedback (left), the pressure and density gradients are parallel along gas streamlines, resulting in no vorticity generation. In the case of strong thermal feedback (right), however, the heating modifies the density gradient inside the planet's luminosity radius (the radius out to which the planet modifies the disc temperature) such that the density and pressure gradients are perpendicular along gas streamlines, sourcing vorticity.}
    \label{fig:cartoon}
\end{figure}

Recently, the thermal feedback from the large accretion luminosity of efficient planet formation that occurs in dust traps has been explored \citep[e.g.][]{cummins22,Pierens2024}. \citet{Pierens2024} showed that the heating from the accreting planet can excite the planet's eccentricity by modifying the disc's torque on the planet \citep[e.g.][]{Chrenko2017,Eklund2017}. Furthermore, \citet{owen17} \textcolor{mod}{proposed that the presence of a hotspot around an accreting planetary embryo should drive vortex formation, as the resulting pressure and density structure drives circulation of gas (see figure~\ref{fig:cartoon}).} The simulations of \citet{cummins22} and \citet{Pierens2024} confirmed that sufficient accretion luminosity could indeed trigger this process and drive vortex formation, with the resulting vortex trapping dust, impacting the planet's subsequent growth. \citet{cummins22} considered the scenario of a planetary core, initially $0.1 M_\oplus$ in mass, situated in a millimetre-bright ``transition disc'' ring \citep{owen_clarke12}, containing $150 M_\oplus$ of dust, and showed that pebble accretion onto the embryo generates substantial accretion luminosity. The simulations demonstrated that anticyclonic vortices were formed once the accretion luminosity could heat the disc beyond the embryo's Hill sphere above its background temperature. They also showed that the presence of a vortex at the planet's location can enhance the pebble accretion rate: being a pressure maximum, the vortex can \textcolor{mod}{provide a larger cross-section for accretion, funnelling pebbles towards the planet, if the radial extent of the vortex exceeds the planet's Hill radius}, creating a new phase of ``vortex-assisted'' pebble accretion. However, the vortex can also act to inhibit the planet's growth when it migrates away from the planet, trapping dust and thus reducing the material available for accretion onto the planet.
\cite{owen17} proposed that the formation of a vortex through this mechanism should result in a lower final planet mass than would be achieved by pebble accretion alone, hypothesising that the vortex would trap the dust and thus prevent it from being accreted onto the embryo, thereby inhibiting the planet's growth. As a vortex is an overdensity in material, it exerts gravitational torques on, and thus exchanges angular momentum with the disc, allowing it to migrate in a similar manner to how a planet does \citep{paardekooper10}. Once the planet is large enough to open a gap in the disc, the (approximate) symmetry of the torques is disrupted, and the vortex may migrate away from the planet. Given a lower rate of vortensity generation, the vortex may separate from the planet earlier and, therefore, accumulate dust before the planet becomes too massive. An important factor in the outcome of the planet's growth, therefore, is how long the vortex remains co-located with the planet, which is set by the dynamical interaction between the vortex and both the planet and the disc.

 The interaction between the planet and vortex can be highly non-linear and dynamic. The size of the vortex can only approach $\sim 4/3H_\ug$ (where $H_\ug$ is the gas pressure scale height) in the radial direction before it begins to launch waves due to the supersonic velocity perturbations \citep[e.g.][]{lyra_lin13}. These waves disturb the disc and may also contribute to the vortex's migration. However, since the presence of a vortex at the planet location increases the pebble accretion rate, which is the source of vortensity generation, this may result in an extended period over which the vortex and planet are co-located: the instability arising from vortex generation, which enhances the pebble accretion rate and thus further strengthens the vortex, has the potential to result in vigorous disc dynamics, where the vortex launches waves while having its energy replenished by the planet's accretion luminosity.

It is, therefore, crucial to investigate the impact of the planet's accretion luminosity across a range of parameter space, not just at the extreme end considered previously. In \cite{cummins22}, planet formation proceeded within a narrow ($\sim 10$~AU) ring at a low background disc temperature (35~K), providing a disc aspect ratio of 0.05 at the planet location; these are conditions conducive to strong thermal feedback from the embryo's accretion luminosity, resulting in vortex formation. \citet{Pierens2024} expanded on these calculations slightly by considering variations in the disc's viscosity and the size of the dust grains (parametrized via their Stokes number), along with considering the planet's migration; however, they parametrized the pebble accretion rather than explicitly resolving it as in \citet{cummins22}. Thus, it is necessary to understand the impact of thermal feedback and vortex formation on planet formation in dust traps to understand its potential under various protoplanetary disc conditions. Therefore, in this work, we expand upon the calculations of \cite{cummins22}, where we use high-resolution simulations which explicitly resolve pebble accretion, to understand the impact of a forming planet's accretion luminosity on its further evolution. 

\section{Observationally motivated dust trap properties}

To establish what constitutes ``typical'' protoplanetary disc conditions for this problem, it is instructive to examine a large study of the distribution of dust in protoplanetary discs with rings; we therefore motivate our setups based on the DSHARP survey \citep{dsharp1}. While this survey specifically targeted bright and, therefore, massive discs and intentionally avoided ``transition'' discs, it is becoming evident that it is these most massive discs that contain pronounced rings \cite[e.g.][]{long18, dsharp1, odisea1, odisea3, kurtovic21}. Thus, such discs are an appropriate starting point, given it is planet formation within such rings that we are concerned with. Note, however, that there is a possible observational bias given the tendency for surveys to focus on bright discs, and brighter discs tend to be larger, making the substructure easier to resolve.

\cite{dsharp6} studied the properties of eight rings across five discs from the DSHARP survey; these specific discs were focussed on for the particularly strong brightness contrasts between their rings, allowing their widths and amplitudes to be clearly defined. Table~\ref{tab:rd_masses} lists the properties of these rings as determined by \cite{dsharp6}. Other notable examples of discs with well-defined, narrow rings include HD~169142 \citep{fedele17, perez19}, HD~97048 \citep{walsh16, vanderplas17}, GM~Aur \citep{maps14}, MWC~480 \citep{long18, maps14}, ISO-Oph~17 and ISO-Oph~54 \citep{odisea3} but these have not been analysed in as much detail on a per-ring basis as those in \cite{dsharp6}. In the ALMA survey presented in \cite{long18}, they estimated ring widths of $\sim 10$ to 30~AU, though at a lower spatial resolution of $\gtrsim 0.1''$ (cf. $\sim 0.035''$ in the DSHARP survey), so it is not clear whether the rings were definitely resolved.

\begin{table*}
	\centering
	\caption[Ringed disc properties]{Disc ring properties from \cite{dsharp6}. Column 1 is the name of the source; column 2 is the number of the ring; column 3 is the distance of the centre of the ring from the central star; column 4 is the width of the ring; column 5 is the estimated midplane disc temperature; column 6 is the aspect ratio of the disc at the ring location; column 7 is the mass of dust in the ring assuming the emission is optically thin, and column 8 is the mass of the dust in the ring correcting for the optical depth.}
	\label{tab:rd_masses}
	\begin{tabular}{lcccccccc}
		\hline
		Source & Ring & $a_0\:/\:$AU & $w\:/\:$AU & $T\:/\:\mathrm{K}$ & $h$ & $M_\ud^\mathrm{thin}/M_\oplus$ & $M_\ud^\mathrm{true}/M_\oplus$\\
		\hline
		AS 209    & 1 &  74 & 3.38 & 15.8 & 0.08 & 27.0 & 31.5\\
		AS 209    & 2 & 120 & 4.11 & 12.4 & 0.09 & 58.7 & 69.8\\
		Elias 24  & 1 &  77 & 4.57 & 22.3 & 0.1  & 35.4 & 40.8\\
        HD 163296 & 1 &  68 & 6.84 & 30.8 & 0.06 & 48.3 & 56.0\\
        HD 163296 & 2 & 100 & 4.67 & 25.3 & 0.07 & 39.0 & 43.6\\
        GW Lup    & 1 &  86 & 4.80 & 10.2 & 0.09 & 33.2 & 37.0\\
        HD 143006 & 1 &  41 & 3.90 & 27.2 & 0.05 &  9.2 &  9.9\\
        HD 143006 & 2 &  65 & 7.31 & 21.6 & 0.06 & 24.0 & 25.6\\
		\hline
	\end{tabular}
\end{table*}

The most significant difference between these discs and the system considered in \cite{cummins22} is that these discs contain multiple concentric rings, with widths of $\sim 5$~AU and dust masses of $\sim 50 M_\oplus$, rather than the millimetre-bright ``transition'' discs with a single ring (i.e. the outer disc) spanning $\gtrsim 10$~AU and containing $\gtrsim 100 M_\oplus$ in solids (it is also worth noting that the wide rings observed in some transition discs have been resolved into multiple rings when observed at higher spatial resolution, e.g. GM~Aur: \citealt{macias18, maps14}; LkCa~15: \citealt{facchini20}). Thus, the rings listed in table~\ref{tab:rd_masses} are more narrow and less massive than those used in the simulations presented in \cite{cummins22}; however, they are further out in the disc, and thus at cooler locations, which bears consequences for the strength of the thermal feedback on the disc.
The aspect ratio ($h$) of the disc can be related to the disc's background temperature ($T_\mathrm{b}$) as a function of radius ($r$) by
\begin{equation}
  h(r) = \sqrt{\frac{k_\mathrm{B} T_\mathrm{b} r}{\mu m_\mathrm{H} G M_*}},
  \label{eqn:aspect_ratio}
\end{equation}
where $k_\mathrm{B}$ is the Boltzmann constant, $\mu$ is the mean molecular weight of the gas, $m_\mathrm{H}$ is the mass of hydrogen, $G$ is the gravitational constant, and $M_*$ is the star's mass. Assuming a background disc temperature profile of $T_\mathrm{b} \propto r^{-1/2}$, the aspect ratio follows $h \propto r^{1/4}$. Thus, while many of the rings listed above are in cold regions of the disc, the disc flaring gives them large gas pressure scale heights. In order for the accretion luminosity to significantly impact the gas dynamics, the luminosity radius must exceed the accreting embryo's Hill radius; the background disc temperature is, therefore, critical when considering the impact of the accretion luminosity. This can be seen in the expression for the ratio of the luminosity radius ($R_\mathrm{L}$, the radius out to which accretion luminosity impacts the disc's temperature, \citealt{rafikov06,owen17}) to the Hill radius ($R_\mathrm{H}$),
\begin{equation}
  \frac{R_\mathrm{L}}{R_\mathrm{H}} = \bigg(\frac{GM_\mathrm{P} \mathit{\Omega}_\mathrm{K} \mathit{\Sigma}_\mathrm{peb}}{8\uppi R_\mathrm{P} \sigma T_\mathrm{b}^4}\bigg)^{1/2},
  \label{eqn:rlrh}
\end{equation}
which depends on the inverse square of the background disc temperature, where $\mathit{\Omega}_\mathrm{K}$ is the Keplerian angular velocity, $\mathit{\Sigma}_\mathrm{peb}$ is the pebble surface density, $R_\mathrm{P}$ is the planet's radius and $\sigma$ is the Stefan-Boltzmann constant. Expressing this instead in terms of the aspect ratio shows that the scaling with the semimajor axis ($a$) is weak, but strong with the inverse of the aspect ratio:
\begin{equation}
  \frac{R_\mathrm{L}}{R_\mathrm{H}} \propto \frac{M_\mathrm{P}^{1/2} \mathit{\Sigma}_\mathrm{peb}^{1/2} a^{1/4}}{R_\mathrm{P}^{1/2} M_*^2 h^4}.
\end{equation}
The accretion luminosity of a forming planet is therefore much more significant in a thinner disc, making it easier to form a vortex.

The rings listed in table~\ref{tab:rd_masses} have temperatures $\sim 30$~K, but aspect ratios between 0.05 and 0.1. \cite{dsharp6} determined the midplane disc temperature via \cite[e.g.][]{chiang_goldreich97}
\begin{equation}
  T_\mathrm{b}(r) = \bigg(\frac{\phi L_*}{8 \uppi r^2 \sigma}\bigg)^{1/4},
\end{equation}
where $\phi=0.02$ is the flaring angle of the disc (i.e. roughly $h \sim \mathrm{atan}\,\phi$), and $L_*$ is the star's bolometric luminosity, allowing an estimate of the disc aspect ratio via equation~(\ref{eqn:aspect_ratio}). By fitting a Gaussian to each of the peaks in the azimuthally averaged intensity profile for each disc and deconvolving the Gaussian beam, \citet{dsharp6} obtained the absolute widths of the rings. They estimated the mass of dust in each ring using the optically thin approximation \cite[e.g.][]{hildebrand83}, but also calculated optical-depth-corrected dust masses.

The dust mass in a ring at a semimajor axis $a$, described by a Gaussian of width (standard deviation) $w$ and amplitude $A$, is given by
\begin{equation}
M_\ud = \int_{r_\mathrm{in}}^{r_\mathrm{out}} 2\uppi r \mathit{\Sigma}_\ud\:\ud r \approx \frac{(2\uppi)^{3/2} A a w}{\kappa_{\nu,\mathrm{abs}} B_\nu(T_\mathrm{b})},
\end{equation}
where $r_\mathrm{in}$ and $r_\mathrm{out}$ enclose the ring radially, $\mathit{\Sigma}_\ud$ is the dust surface density and $B_\nu$ is the Planck function. The dominant source of uncertainty in this dust mass estimate is the opacity, $\kappa_{\nu,\mathrm{abs}}$, which depends on the maximum grain size assumed, the grain size distribution, as well as their composition and shape, and could be uncertain by a factor of 10 \citep{dsharp5}. \cite{dsharp6} adopted a grain maximum grain size ($s_\mathrm{ max}$) of 1~mm based on the width of the rings and an estimate of the dust's Stokes number and thus used an opacity of 2~cm$^2$~g$^{-1}$, given by the opacity model presented in \cite{dsharp5}. However, if we instead assume that the maximum grain size is given by the fragmentation limit ($s_\mathrm{frag}$) -- since a dust population trapped in a pressure maximum should reach fragmentation-coagulation equilibrium -- the dust masses will be modified by a factor $\kappa_{\nu,\mathrm{abs}}(s_\mathrm{max} = 1\,\mathrm{mm})/\kappa_{\nu,\mathrm{abs}}(s_\mathrm{max} = s_\mathrm{frag})$.

Determining the fragmentation-limited grain size requires an estimate of the gas density, which is difficult to do observationally due to the lack of optically thin tracers \citep[e.g.][]{Williams2014,Bergin2018}. \cite{dsharp6} placed upper and lower boundaries on the gas surface density ($\mathit{\Sigma}_\ug$) based on gravitational stability and dust-to-gas ratio arguments respectively, resulting in typical values of $\mathit{\Sigma}_\ug$ between $\sim 0.1$ and 10~g~cm$^{-2}$. Similarly, an estimate of the strength of turbulence is required; again, while this is difficult to constrain, a typically adopted value is $\alpha \sim 10^{-4}$. Thus, the maximum grain size, limited by fragmentation, can be expressed as \cite[e.g.][]{birnstiel09}:
\begin{equation}
  \begin{split}
    s_\mathrm{frag} & \approx \frac{2\mathit{\Sigma}_\ug}{\uppi\alpha\rho_\bullet}\frac{v_\mathrm{frag}^2}{GM_*}\frac{r}{h^2} \\
    & \approx 1\:\mathrm{cm} \: \bigg(\frac{\mathit{\Sigma}_\ug}{1\:\mathrm{g\:cm}^{-2}}\bigg) \bigg(\frac{\alpha}{10^{-4}}\bigg)^{-1} \bigg(\frac{M_*}{2M_\odot}\bigg)^{-1}\bigg(\frac{r}{80\:\mathrm{AU}}\bigg)
    \\
    & \:\:\:\:\: \times \: \bigg(\frac{h}{0.1}\bigg)^{-2} \bigg(\frac{v_\mathrm{frag}}{10\:\mathrm{m\:s}^{-1}}\bigg)^2.
  \end{split}
\end{equation}
where $\rho_\bullet$ is the internal density of the dust grains. Since we are considering regions of the disc beyond the water iceline, grains are assumed to be icy and thus have fragmentation velocities of $v_\mathrm{frag} \sim 10$~m~s$^{-1}$, which permits maximum grain sizes $s_\mathrm{frag} \sim 1$~cm. Using the \cite{dsharp5} opacity model, $\kappa_{\nu,\mathrm{abs}}(s_\mathrm{max} = 1\,\mathrm{cm}) \sim 0.5$~cm$^2$~g$^{-1}$, suggesting that the dust mass in these rings may well be up to a factor $\sim 4$ higher than those reported when assuming a maximum grain size of 1~mm. Note that this is comparable to the upper bound of the uncertainty ranges calculated by \cite{dsharp6} and more consistent with those of \cite{Stadler2022}, who argue that some dust ring masses were underestimated in earlier work. While there is uncertainty in the exact masses in these rings, the values observed from observations can be used as a guide and supplemented by theoretical considerations. Thus, it is these supplemented values that we consider in this work.

One important question to answer is how much of the mass in the dust trap the planet will accrete. Many of the observed rings contain much of their mass in large grains, and rings seem to be ubiquitous in discs observed in high spatial resolution observations.
\cite{cummins22} showed that, both with and without the thermal feedback from its accretion luminosity, a planet could grow beyond the classical pebble isolation mass in a dust trap. Thus, if planet formation is almost inevitable in these rings, as previously argued, one should expect to see many planets with large metal masses, i.e. planets with solid masses well above the typical mass required for runaway gas accretion, if they can accrete a large fraction of the dust within the ring. Therefore, it is of interest to determine how planet formation efficiency in dust traps varies with both the aspect ratio, upon which the classical isolation mass depends, as well as the accretion rate set by the dust mass. The latter may be important if the isolation mass can be exceeded simply because the accretion timescale is shorter than the gap opening timescale.

Following \cite{lin_papaloizou86}, the gap-opening timescale at a distance $H_\ug$ from the planet can be estimated from $\mathit{\Sigma}_\ug/\dot{\mathit{\Sigma}}_\ug$, with $\dot{\mathit{\Sigma}}_\ug$ the accretion rate, which gives
\begin{equation}
  \begin{split}
    t_\mathrm{gap} & \approx \bigg(\frac{M_\mathrm{P}}{M_*}\bigg)^{-2} \bigg(\frac{H_\mathrm{g}}{a}\bigg)^{5} \mathit{\Omega}_\mathrm{K}^{-1} \\
    & \sim 8\:\mathrm{kyr} \: \bigg(\frac{M_\mathrm{P}}{20M_\oplus}\bigg)^{-2} \bigg(\frac{a}{35\:\mathrm{AU}}\bigg)^{3/2} \bigg(\frac{h}{0.05}\bigg)^{5} \bigg(\frac{M_*}{2M_\odot}\bigg)^{3/2}
  \end{split}
\end{equation}
for a typical pebble isolation mass of $20M_\oplus$.
This is longer than the growth timescale of the planet seen in \cite{cummins22} (see their figure~3) where, without considering the accretion luminosity (since the waves launched by both the temperature disturbance and the vortex would also contribute to gap formation), the planet continued to accrete from $\sim 20M_\oplus$ at 5~kyr to $\sim 40M_\oplus$ at 10~kyr, at which point pebble accretion ceased.
The timescale for halting pebble accretion in the classical scenario will be shorter, however, as a pressure bump will form exterior to the planet's orbit, halting pebble drift, before the co-orbital region is significantly depleted of material, thus the above cannot be used to accurately determine the pebble isolation mass. This was also argued by \cite{bitsch18} in their investigation into the pebble isolation mass, and they showed that pebble drift is halted when gap depths are only around 10--20\%. This can be seen in figure~3 of \cite{cummins22}, where pebble accretion is halted despite the planet only having created a shallow gap.

Another issue to address is whether this mechanism can really produce large-scale vortices, i.e. comparable in size to those inferred from the observations of IRS~48 and HD~142527, as found by \cite{owen17}. It is worth noting that in their simulations they used a disc aspect ratio of 0.1 at the planet location. Since discs with a larger scale height can support vortices with a larger radial extent \citep[e.g.][]{lyra_lin13}, their simulations produced an asymmetry similar in size to those produced by other mechanisms and similar to the scales inferred from observations. However, their vortensity generation was not calculated self-consistently, and the planet was kept at a low mass, which allowed multiple small-scale vortices to form and eventually merge to form one large vortex.
Thus, it is not \textit{a priori} obvious how this will vary with different dust masses and thus different accretion rates. The strength of the generated vortex can be approximated by ${\rm D}\omega/{\rm D}t \Delta t$; thus, while a high accretion rate may generate vorticity quickly, it will produce a high-mass planet, which will quickly create a gap, preventing further accretion and therefore, vorticity generation. Similarly, the highly dynamic interaction between the vortex and the planet may lead to the vortex separating from the planet quickly, preventing any further vorticity generation.
However, a lower initial pebble accretion rate would allow vorticity to be generated more slowly over a longer period of time and may result in a stronger, larger vortex. Alternatively, a reduced vortex generation rate may permit the formation of multiple small-scale vortices, each of which separate from the planet and eventually merge into one large-scale vortex.

In addition to the effect of accretion luminosity enhancing the pebble accretion rate, some additional outcomes of the simulations in \cite{cummins22} can be explored further by performing simulations across a wide range of parameter space. For example, those simulations showed that since the dust populations most efficiently accreted onto the embryo were also those most rapidly trapped in the vortex, meaning an inversion of the size distribution of pebbles was produced with smaller, rather than larger grains containing more of the mass. We therefore wish to see whether this is a general outcome of this process or whether it only occurs in the most extreme of cases. This could potentially be an observable indicator that this mechanism has taken place: if vortices are short-lived and we \textit{are} seeing them shortly after formation, then the dust population within the residual ring(s) may not have had time to re-equilibrate into a growth-fragmentation equilibrium distribution \citep{birnstiel11}.

Another outcome we wish to investigate further is how the vortex-assisted pebble accretion rate varies with the scale height of the disc. As discussed in \cite{cummins22}, the vortex-assisted pebble accretion rate can be interpreted as the capture rate of pebbles by the vortex, which, therefore, depends on the vortex's radial extent. The accretion rate is enhanced by a factor $2H_\ug^2/3R_\mathrm{H}^2$ and thus depends on the aspect ratio of the disc; the explicit dependence on the aspect ratio will therefore allow us to determine whether this relation holds across parameter space.

Even if it is insufficient to trigger vortex formation, the accretion luminosity may affect the accretion rate simply by altering the velocity streamlines around the planet \citep[e.g.][]{Pierens2024}. Thus, there may be a narrow region of parameter space where thermal feedback is insufficient to produce vortex-assisted pebble accretion but can still affect the accretion rate, ultimately resulting in a difference in planet mass. We therefore want to determine how the accretion luminosity of a pebble-accreting planetary embryo affects its growth in a general sense, rather than just its ability to source vorticity and generate a vortex.

First, we provide details of the dust masses and disc aspect ratios for our parameter study and recap the details of the hydrodynamics simulations performed by \citet{cummins22}. We then present the results of simulations, focussing on the final planet mass reached and the consequences for the evolution of the disc for the various disc parameters used. We then discuss what these results imply about planet formation in dust rings, both in terms of the planet's growth and observable disc substructure.

\section{Simulation Setup}
To investigate the impact of an accreting embryo's luminosity on its growth across a range of disc properties, we performed global 2D hydrodynamics simulations of a protoplanetary disc containing a dust ring with a planetary embryo accreting dust from within it. Our simulation framework is identical to that described in \citet{cummins22}: we use our modified version of \texttt{FARGO3D} \citep{fargo3d}, which includes a prescription for resolving pebble accretion as well as the associated thermal feedback on the gas. The pebble accretion procedure is tested and described in detail in \citet{cummins22}; here, we simply describe our disc setups.

We used a central star with mass $M_* = 2 M_\odot$, hosting a disc of mass $M_\mathrm{disc} = 0.05 M_*$, 175~AU in radial extent and with a constant alpha viscosity of $\alpha = 10^{-4}$. The gas surface density is initialised according to
\begin{equation}
  \mathit{\Sigma}_\ug = \frac{\mathit{\Sigma}_0}{2}\bigg(\frac{r}{R_0}\bigg)^{-1}\bigg[1+\mathrm{erf}\bigg(\frac{r-R_1}{\sqrt{2}\sigma}\bigg)+\epsilon\bigg],
  \label{eqn:sigma_g_c4}
\end{equation}
with $R_1 = 1.2R_0$, $\sigma = 0.663R_0$ and $\epsilon = 10^{-3}$. We chose $R_0 = 35$~AU, and $\mathit{\Sigma}_0 \approx 14.5$~g~cm$^{-2}$ to set the total disc mass. The gas pressure scale height is given by
\begin{equation}
  H_\mathrm{g}(r) = h_0 R_0 \bigg(\frac{r}{R_0}\bigg)^{5/4}.
  \label{eqn:gasscaleheight}
\end{equation}
We used a local isothermal equation of state, which maximises the vertically integrated gas pressure at $r \approx R_0$ with these gas surface density and pressure scale height profiles.

We used four dust species, with sizes 1~$\upmu$m, 1~mm, 3.16~mm and 1~cm in an MRN size distribution, each with densities of 3~g~cm$^{-3}$. The initial dust density distribution for each dust species was calculated by solving the steady-state advection-diffusion equation\textcolor{mod}{, which yields full-width at half-maximum (FWHM) ring widths of approximately 3~AU, 4~AU and 7~AU for the 1~cm, 3.16~mm and 1~mm dust grains respectively}. We varied the initial mass in the dust trap from $50 M_\oplus$ to $150 M_\oplus$ in steps of $25 M_\oplus$.

Since the gas surface density is the same for each simulation, the dynamical properties of the dust grains, i.e. their Stokes numbers, are preserved. Similarly, while the dust mass is varied between simulations, the initial gas mass is the same; the results from tests presented in \citet{cummins22} showing that the initial gas distribution is stable both against self-gravity and the Rossby wave instability therefore remain valid. \textcolor{mod}{Additionally, we confirmed that the initial dust ring is stable against the dusty Rossby wave instability \citep[e.g.][]{Liu2023} by performing a simulation for 1000 orbits without a planetary embryo.}

We also varied the disc temperature at each dust mass by using an aspect ratio $h(a)$ of 0.035, 0.05 and 0.1. While an aspect ratio of $h(a) = 0.1$ requires a relatively high temperature at the planet's semimajor axis, the scale height determines the impact of the thermal feedback.

In each simulation, we used an initial embryo mass of $M_\mathrm{P} = 0.1 M_\oplus$, positioned at the centre of the dust trap. To isolate the physics of the problem and to allow comparison with the results of \citet{cummins22}, we do not include planet migration at this stage. The embryo's potential is smoothed according to
\begin{equation}
  \mathit{\Phi}_\mathrm{P}(d) = -\frac{GM_\mathrm{P}}{(d^4+\epsilon^4)^{1/4}},
  \label{eqn:potsmooth}
\end{equation}
where $d$ is the distance from the embryo, with $\epsilon = 0.35 R_0$, which accurately reproduces the potential down to within the embryo's Hill sphere while preventing a singularity at the embryo's location. The accretion luminosity was calculated using
\begin{equation}
  L_\mathrm{P} = \frac{GM_\mathrm{P} \dot{M}_\mathrm{peb}}{R_\mathrm{P}},
  \label{eqn:L_accr}
\end{equation}
with the planet radius, $R_\mathrm{P}$, calculated using the mass-radius relationship of \cite{fortney07}, with an ice mass fraction of 0.5. The hotspot temperature was calculated using
\begin{equation}
  T = T_\mathrm{b}\bigg(1 + \frac{R_\mathrm{L}^2}{d^2+s_\mathrm{T}^2}\bigg)^{1/4},
  \label{eqn:disctemp}
\end{equation}
with a luminosity radius $R_L = \sqrt{L_\mathrm{P}/(16\uppi\sigma T_b^4(a))}$ \citep{owen17}. We use a temperature smoothing length of $s_\mathrm{T}=0.65 R_{\mathrm{H},0}$, where $R_{\mathrm{H},0}$ is the Hill radius of the planet at its initial mass. Thermal feedback is implemented as described in \citet{cummins22}, where the temperature of the disc is calculated \textcolor{mod} {using Equation~\ref{eqn:disctemp}, based on the current accretion rate of the planet. This temperature profile is used to update the gas sound speed every time-step.}
We used a resolution of 1920 cells in radius and 7680 in azimuth, spanning $r \in [0.54, 5.0] R_0$ and $\phi \in [0, 2\uppi]$. The resolution is maximised at the planet location by using variable cell sizes in radius, defined by
\begin{equation}
  \Delta r = c_1 r\bigg\{1-c_2\exp\bigg[-\bigg(\frac{r-a}{\sqrt{2}w}\bigg)^2\bigg]\bigg\}.
  \label{eqn:cell_spacing}
\end{equation}
Thus, 660 cells cover $r \in [0.8, 1.2] a$. \citet{cummins22} demonstrated this grid setup and resolution was sufficient to resolve the pebble-accretion and vortex dynamics. 

We used an open boundary condition at the inner edge of the disc, allowing the material to leave the computational domain. In contrast, the dust and gas radial velocities are reflected at the outer boundary. The dust density is set to the floor value beyond the outer boundary, and the gas surface density is set by equation~(\ref{eqn:sigma_g_c4}). Wave-killing zones were employed near the boundaries to suppress any artefacts that can arise from using a finite domain. These extend from the inner (-) and outer (+) boundaries to $r_\mathrm{damp}^{\pm} = r_\mathrm{bound}^{\pm} 1.15^{\mp 2/3}$, with the damping implemented as described in \cite{devalborro06}, using a damping timescale of 1.5 orbital periods at the respective boundary radii.

\section{Results}
We present the results of simulations in which the initial mass within the dust ring was varied between $50 M_\oplus$ and $150 M_\oplus$, for aspect ratios of $h(a) = 0.035$, 0.05 and 0.1, first showing the impact on the final planet masses achieved, before focussing on the evolution of the discs and their substructure. Note that the $150 M_\oplus$, $h(a) = 0.05$ simulation is the same one from \cite{cummins22}, which we include here for completeness in our parameter study.

\subsection{Planet growth}
\subsubsection{Final planet mass}
\begin{figure}
  \centering
  \includegraphics{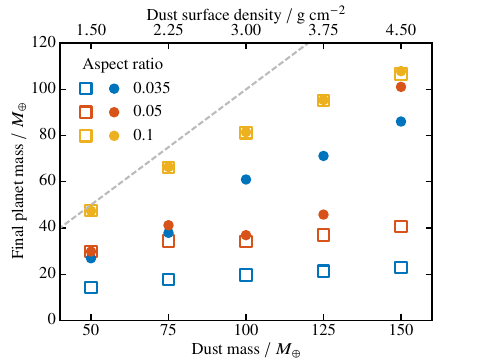}
  \caption[Final planet mass reached in dust rings of different masses and temperatures]{Final planet mass for each ring mass and aspect ratio. Filled circles are for simulations with thermal feedback from pebble accretion; open squares are for simulations without. The grey dashed line shows the mass the planet would reach if it accreted all dust within the ring. Also shown is the peak dust surface density at the location of the planet, which scales linearly with the initial dust mass.}
  \label{fig:paramstudy_summary}
\end{figure}

Figure~\ref{fig:paramstudy_summary} shows the mass that the embryo reached via pebble accretion after 60~kyr ($\sim 400$ orbits) within each dust ring, as a function of the dust mass within the ring and the disc aspect ratio at the planet location with and without including the thermal feedback on the disc. In each simulation, the planet reaches its final mass within 60~kyr.

The final planet's mass generally increases with both the disc's scale height and the ring's initial mass. The thermal feedback results in a final mass that is always greater than or equal to that reached when neglecting this effect. This is most apparent in the coldest discs, where there is a factor of $\approx 2$ increase in the least massive ring, up to a factor $\approx 4$ increase in the most massive ring. 
An increase in the final planet mass with increased scale height is seen for growth with and without thermal feedback on the disc.

In the $h(a) = 0.05$ discs, there is a non-monotonic increase in planet mass with increasing ring mass: the $100 M_\oplus$ ring produces a lower mass planet than the $75 M_\oplus$ ring; otherwise, the planet mass increases with ring mass. Similarly, the increase in final planet mass due to thermal feedback increases non-monotonically with increasing ring mass, with no difference in the least massive ring and up to a factor of 2.5 in the most massive ring.
In the discs with the highest aspect ratio, thermal feedback only affects the planet's growth in the most massive ring, with an increase of $1.2 M_\oplus$ ($\approx 1\%$) over that reached when neglecting accretion luminosity; otherwise, there is no discernible difference when including the thermal feedback. Despite the embryo's accretion luminosity having little-to-no effect on its growth, its final mass increases approximately linearly with the ring mass -- a stronger scaling than one might naively expect given the results at lower aspect ratios.

\subsubsection{Accretion rates}
\begin{figure}
  \centering
  \includegraphics{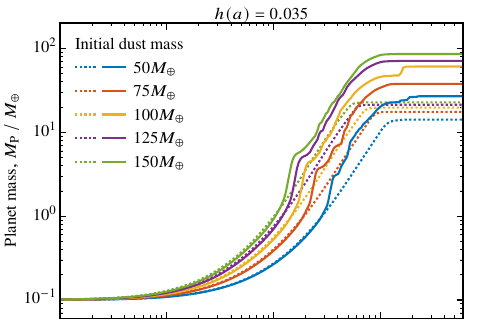}
  \includegraphics{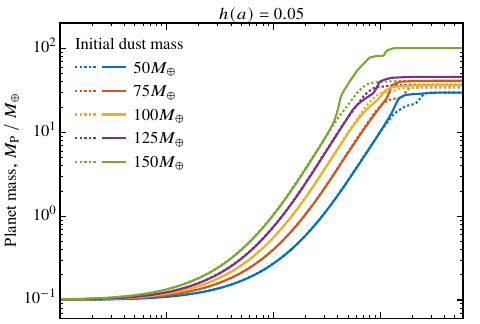}
  \includegraphics{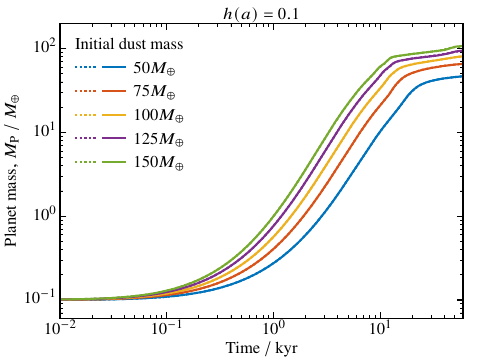}
  \caption[Planet mass comparison]{Planet mass as a function of time for a range of initial dust masses and disc temperatures. Each panel compares the planet's growth for different dust masses in a dust trap with the same scale height. The dashed lines show the planet's growth without accounting for the thermal feedback on the disc from its accretion luminosity, while the solid lines show the growth when accounting for this effect.}
  \label{fig:mp_h_nh_fixedh}
\end{figure}

How the planet's final mass varies with the initial dust mass and aspect ratio can be better understood by considering the pebble accretion rates over time. Figure~\ref{fig:mp_h_nh_fixedh} shows the growth of the planet over 60~kyr for each ring mass and aspect ratio. For the lowest aspect ratio, and thus the lowest background disc temperature, the impact of the thermal feedback is clearly apparent for all initial ring masses. A distinct jump in accretion rate is seen after $\sim 1$~kyr, at progressively later times for lower dust masses ($\sim 3$~kyr for the lowest dust mass). In each case, the strong thermal feedback results in the formation of a vortex, which enhances the pebble accretion rate, producing a significantly more massive planet than when neglecting the thermal feedback.

\begin{figure}
  \centering
  \includegraphics{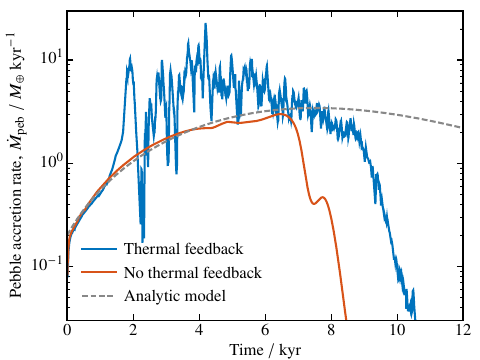}
  \caption[Pebble accretion rates for $100 M_\oplus$, $h(a) = 0.035$.]{Pebble accretion rates onto the embryo, with and without thermal feedback, in the $100 M_\oplus$, $h(a) = 0.035$ ring. The blue line shows the pebble accretion rate in the simulation with thermal feedback. The red line shows the pebble accretion rate in the simulation without thermal feedback. The grey dashed line shows the pebble accretion rate from a simple analytic calculation, which does not account for gap-opening or the embryo's accretion luminosity.}
  \label{fig:par_100_0035}
\end{figure}

The pebble accretion rates show similar behaviour for each dust mass in the $h = 0.035$ rings, closely following the accretion rate when neglecting thermal feedback until vortices form, at which point it rapidly increases. While the vortex is co-located with the planet, the pebble accretion rate, although showing significant fluctuations, has a reasonably steady time-averaged value, as was seen in \cite{cummins22}. The pebble accretion rate then begins to decay after $\sim 5$ to $9$~kyr, depending on (but not necessarily scaling with) the ring mass, before rapidly dropping off once the vortex separates from the planet. This can be seen in figure~\ref{fig:par_100_0035}, which shows the pebble accretion rate over the first 12~kyr onto the planet accreting in the $100 M_\oplus$ ring. This is largely representative of the pebble accretion rates for each ring mass in the $h(a) = 0.035$ discs; however, the increase in accretion rate due to the formation of the vortex begins earlier, and the pebble accretion rate is higher, with increasing ring mass. Furthermore, only in the $50 M_\oplus$ case does the accretion rate sharply drop off \textit{before} it does in the simulation without thermal feedback -- this is likely to be because the drop-off in accretion rate without thermal feedback is shallower for lower dust masses, as the planet mass and therefore pebble accretion rate is lower.

The simulations for the intermediate scale height with $h(a) = 0.05$ appear to show a range of behaviours, suggesting that they may cross the region of parameter space over which the thermal feedback becomes important for the planet's growth. As discussed in \cite{cummins22}, the $150 M_\oplus$ dust trap provides an accretion rate sufficient to result in vortex formation and, therefore, a phase of vortex-assisted pebble accretion. For the $125 M_\oplus$ ring, however, an initial enhancement in the accretion rate with respect to the case without thermal feedback is seen, followed by a reduction in accretion between 6 and 9~kyr, before the accretion rate increases again. This does not follow the same course of events as in the $150 M_\oplus$ ring, purely because the rate of vortensity generation is reduced and the vortex that forms is not large enough to enhance the accretion rate. 
Similar behaviour is seen at $100 M_\oplus$, although without the initial increase in accretion rate, due to even weaker vortex formation.

Moving to even lower dust masses, the behaviour of the accretion rate and, thus, the planet's growth over time changes further still. At $75 M_\oplus$ there is an enhancement in accretion rate after 10~kyr and the planet ends up more massive than in the $100 M_\oplus$ case -- this is the only instance of planet growth curves crossing; incidentally, the final planet mass in the simulation without thermal feedback is also higher in the $75 M_\oplus$ ring than the $100 M_\oplus$ ring, suggesting that there must be another effect at play here. We speculate this is because, in these simulations, the accretion timescale and gap-opening timescales are comparable, allowing the embryo to continue accreting. At the same time, it opens a gap, and as such, different physical processes are responsible for slowing accretion when transitioning from an accretion-dominated to a gap-opening-dominated shut-off regime. For the lowest dust mass of $50 M_\oplus$, similar behaviour is observed, with an increase in accretion rate with respect to the simulation without thermal feedback, but the planet's final mass is the same.

\begin{figure}
  \centering
  \includegraphics{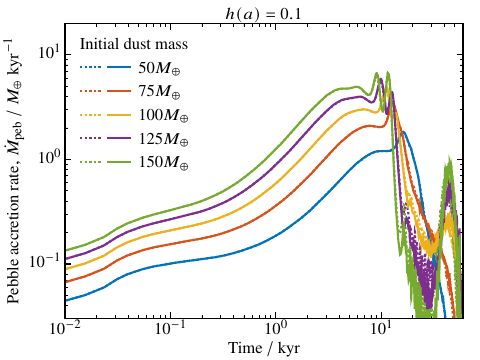}
  \caption[Pebble accretion rate comparison for largest scale height disc.]{Pebble accretion rate as a function of time for a range of initial ring masses for the disc with the largest scale height. The dashed lines show the accretion rate without accounting for the thermal feedback on the disc from its accretion luminosity, while the solid lines show the accretion rate when accounting for this effect.}
  \label{fig:mpebdot_h_nh_fixedh}
\end{figure}

For the largest scale height, the thermal feedback from the embryo's accretion luminosity is negligible, and as such, there are no significant differences between the planet's growth with and without including its accretion luminosity. This can also be seen in figure~\ref{fig:mpebdot_h_nh_fixedh}, which shows the pebble accretion rates for this aspect ratio. The accretion rates with and without thermal feedback are indistinguishable for each ring mass until around 10~kyr, where slight departures can be seen in the most massive rings. Given that the pebble accretion rate scales with the square of the planet's Hill radius, it is no surprise that slight differences in accretion rate become amplified at later times for the most massive planets. However, as shown in figure~\ref{fig:mp_h_nh_fixedh}, this has little effect on the planet's final mass.

An interesting feature is the bump in pebble accretion rate onto the planets in the $100 M_\oplus$, $125 M_\oplus$ and $150 M_\oplus$ rings with $h = 0.1$ between 30--55~kyr. Only in these rings, at this scale height, does the planet become massive enough to create a pressure bump outside of its orbit -- albeit weak -- which concentrates the pebbles into a ring. As the planet keeps on growing during this gap-opening phase, however, its Hill sphere extends into this ring and thus accretes from it.

Weak vortices do in fact form in the most massive rings with $h=0.1$, but quickly separate from the planet and thus have a negligible impact on the evolution of the disc or the planet's growth; the final planet mass is set purely by planet-disc interactions. A more detailed look at the vortex formation process and the conditions under which the vortex affects the accretion rate are given in \textsection\ref{sec:vapa}. The fluctuations in pebble accretion rate onto the most massive planets seen beyond $\sim 20$~kyr in figure~\ref{fig:mpebdot_h_nh_fixedh} are due to the accumulation of dust around L$_5$; the planet does not reach a high enough mass to carve a gap in the disc. Thus, as the smaller (millimetre-sized) dust grains are less strongly concentrated around L$_5$ they can enter the planet's Hill sphere as a vortex attracts them on passing. \textcolor{mod}{Given that this happens in the simulations both with and without thermal feedback, this vortex cannot have formed via the embryo's accretion luminosity and thus likely forms due to the planet's spiral wake: the shock generated by the planet is a source of vorticity \cite[e.g.][]{dong11, devalborro07, lin_papaloizou10}, which is stronger in the leading shock, once the planet opens a gap, the small vortex structure survives until dissipation.}

\subsubsection{Quantifying the thermal feedback from accretion luminosity}
\begin{figure}
  \centering
  \includegraphics{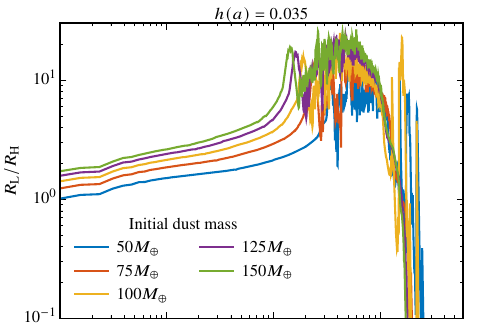}
  \includegraphics{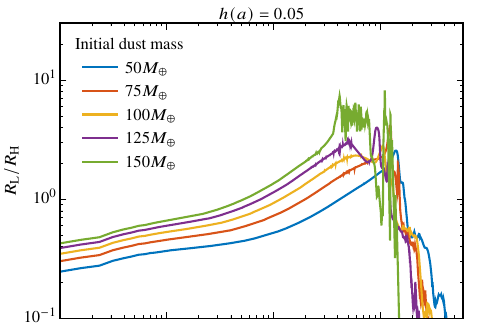}
  \includegraphics{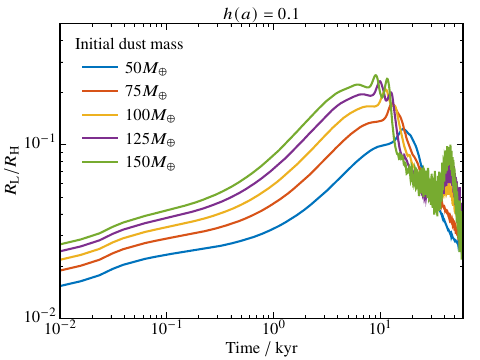}
  \caption[Luminosity radius to Hill radius ratios]{Ratio of the accreting embryo's luminosity radius to its Hill radius as a function of time for a range of initial dust masses. Each panel compares this ratio for different dust masses in a dust trap with the same aspect ratio indicated above each panel.}
  \label{fig:rlrh_fixedh}
\end{figure}

The planet's accretion luminosity can only significantly affect the disc if its luminosity radius exceeds its Hill radius \citep{owen17}. The impact of the accretion luminosity on the planet's growth can therefore be understood by considering how the ratio of these two quantities varies between simulations. Figure~\ref{fig:rlrh_fixedh} shows the time evolution of this ratio for each initial dust mass and gas pressure scale height. This ratio is proportional to the square root of the pebble accretion rate and thus increases with the mass in the ring. Similarly, it is inversely proportional to the square of the background disc temperature and thus decreases with increasing aspect ratio.

The luminosity radius, being the distance from the planet at which the temperature due to its accretion luminosity equals that due to the stellar luminosity alone, is smaller for hotter discs (cf. equation~(\ref{eqn:rlrh})).
For the coldest disc, the embryo's luminosity radius exceeds its Hill radius from the beginning of the simulation for all dust masses, resulting in strong thermal feedback and vortex formation until the vortex separates from the planet, at which point the ratio falls below unity. 
For the disc with $h(a) = 0.05$, the luminosity radius is initially smaller than the Hill radius. Still, as the embryo grows, its Hill sphere expands, enabling it to accrete from a greater distance. This increases its accretion rate, and this ratio eventually exceeds unity for all ring masses, at times ranging from $\sim 500$~yr in the most massive ring to $\sim 4$~kyr in the least massive. In the $150 M_\oplus$ ring, this results in vortex formation and thus an enhancement in accretion rate at around 3.5~kyr, once this ratio exceeds $\approx 3.5$. The luminosity to Hill radius ratio also reaches $\approx 3$ in the $125 M_\oplus$ ring, and a vortex does form. However, it quickly separates from the planet, preventing a phase of vortex-assisted pebble accretion from occurring. Vortex formation does not occur in the least massive rings at this aspect ratio.
The luminosity radius never exceeds the Hill radius in the disc with the largest aspect ratio. Their ratio's maximum value decreases with ring mass, from $\approx 0.25$ in the most massive ring to $\approx 0.1$ in the least massive.

\subsection{Disc evolution}
In addition to its effect on the planet's growth, the thermal feedback on the disc has consequences for the evolution and structure of the disc, with the formation of vortices producing non-axisymmetric features as well as the axisymmetric substructure formed as the planet carves a gap in the initial ring. The formation of substructure also affects the size distribution of dust grains, as the efficiency of trapping in pressure maxima is grain-size-dependent.

\subsubsection{Dust distribution}
In the disc with the largest scale height, the only dust remaining once pebble accretion has ceased is that in the co-orbital region. Only in the $150 M_\oplus$ ring does the planet open a partial gap and create a ring outside of its orbit, but, as described above, the planet can accrete the dust from this ring, making it a short-lived feature. The vortices that form in the most massive rings are weak and therefore don't trap significant amounts of dust, hence the remaining dust gradually settles towards L$_5$.
For the $h(a) = 0.05$ discs, regardless of ring mass, the remaining dust resides in the new axisymmetric pressure maxima that form either side of the planet's orbit. However, the largest dust grains are not present in the inner ring -- they are only found in the outer ring and in L$_5$. The millimetre-sized grains are found in both rings and around L$_5$.
In the coldest rings, all remaining dust is trapped in both rings, enclosing the planet's orbit and L$_5$.

The dust remaining in the co-orbital region therefore accumulates around the L$_5$ Lagrange point for all aspect ratios, even in the least massive rings. This accumulation in L$_5$ is expected from theory. As discussed in \citet{cummins22}, trapping in L$_5$, rather than L$_4$ has been seen previously \citep[e.g.][]{rodenkirch21}. This result is explained by gas-drag destabilising dust at L$_4$ while modifying the location of dust's stable point around L$_5$ \citep{murray94}. This trapping at L$_5$ is particularly important concerning the hypothesis put forward by \cite{owen17}, that dust released by the vortex upon its eventual dissipation would re-form a ring, restarting pebble accretion. In \cite{cummins22}, we speculated that perhaps, if the final planet mass is not too high (i.e. below the gap-opening mass), then the remaining dust would not accumulate in L$_5$ but re-form a ring. However, the results of these simulations suggest that even in the least massive rings, at this viscosity, the planets produced are still massive enough ($\sim 10$--$20 M_\oplus$), to break the azimuthal symmetry of the ring, leaving dust to settle at L$_5$.

\begin{figure*}
  \centering
  \includegraphics{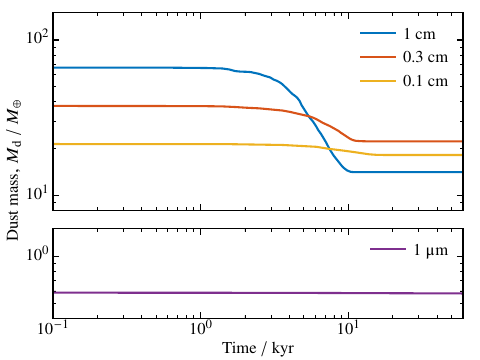}
  \includegraphics{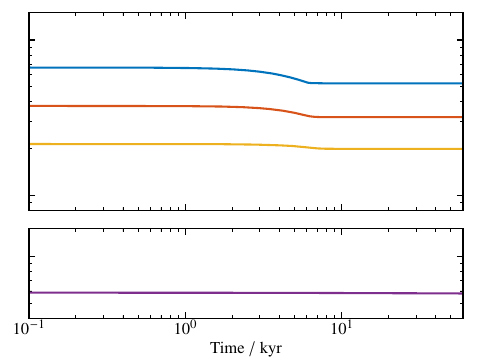}
  \caption[Impact of thermal feedback on the dust size distribution]{Mass in each dust species over the simulation duration, for the $125 M_\oplus$, $h = 0.035$ ring. The left and right-hand panels are for the simulation with and without thermal feedback respectively. When vortex formation increases the pebble accretion rate onto the embryo, the centimetre-sized grains are both preferentially attracted towards the vortex and accreted onto the embryo.}
  \label{fig:dustmass_0035_125}
\end{figure*}

In all simulations, the larger dust grain populations lose the largest fractions of their mass. The $150 M_\oplus$ ring with an aspect ratio of 0.05 results in an inversion of the size distribution between the millimetre and centimetre grains, as described in \cite{cummins22}. However, this is the only instance in which this occurs and is not a general outcome.
Because these largest grains have Stokes numbers closest to unity, they are preferentially accreted onto the embryo. The thermal feedback on the disc can act to emphasise this effect: as higher pebble accretion rates generally result in stronger thermal feedback, then if a vortex forms, this increases the pebble accretion rate and reduces the mass in the largest dust species as they drift towards the vortex the fastest. An example of this is shown in figure~\ref{fig:dustmass_0035_125}, which shows the mass in each dust species in the $125 M_\oplus$, $h = 0.035$ ring, with and without thermal feedback. This illustrates how the centimetre-sized grains lose the largest fraction of their mass without the size distribution being inverted.

In the $h = 0.1$ rings, no inversion of the size distribution results from the initial pebble accretion phase. However, as the planet forms a weak pressure maximum outside of its orbit, only the highest Stokes number grains are trapped there. In contrast, the millimetre grains remain in the co-orbital region, destined for L$_5$. Thus, once the planet accretes from the resulting ring (as described above), the reservoir of centimetre grains is depleted, and the millimetre grains dominate the final disc mass.
This suggests that the formation of pressure maxima of any kind by the planet helps preserve the largest dust grains. While these are accreted onto the planet most efficiently, unless the planet accretes all of the dust before opening a gap, the rest ends up in the vortex or the resulting rings. If trapped in a vortex, they end up in L$_5$ once it dissipates; if trapped in rings, they remain there. If no pressure maxima forms or only weak maxima form, such as those formed by the planet in the $h = 0.1$ rings, the population of centimetre-sized grains is almost entirely depleted.

\subsubsection{Vortex lifetimes}
The lifetimes of the vortices formed from the thermal feedback on the disc range from $\sim 250$ to 450 orbits, with a tendency towards longer lifetimes in lower mass rings. We measured the vortex lifetime as the time between formation and the point at which the vortensity reduces to the background level, determined by eye. As vortices only form consistently via this process in the $h = 0.035$ rings, a comparison as a function of ring mass can only be made for this aspect ratio.

In the most massive ring, the vortex survives for 250 orbits. The vortex lifetime increases to around 300 and 450 orbits in the $125 M_\oplus$ and $100 M_\oplus$ rings respectively.
In the $75 M_\oplus$ ring, there are two significant vortices that form: the initial vortex that forms around the planet separates from the planet after $\approx 40$ orbits, but the embryo continues to accrete, sourcing vortensity resulting in a second vortex, which separates after 110 orbits. Both vortices dissipate after a further $\sim 200$ orbits. A similar outcome is seen in the least massive ring, although these survive for $\sim 400$ orbits.

The three main effects that can weaken the vortices are viscosity, shocks, and increased shear, which the vortices experience during horseshoe turns. Since the viscosity is relatively low, shocks and the shear during horseshoe turns are the dominant effects. We suspect this can explain why the vortex lifetime in the $75 M_\oplus$ ring is an outlier from the trend: there are two vortices on horseshoe orbits, each generating spiral shocks, which are therefore constantly disrupting each other. This is also the case in the $50 M_\oplus$ ring, however, as the planet is less massive than in the $75 M_\oplus$ ring, the horseshoe region is narrower, thus there is less shear, allowing the vortices to survive longer. In the $h = 0.05$, $150 M_\oplus$ ring, the vortex survives for $\sim 500$ orbits, approximately twice as long as in the corresponding $h = 0.035$ ring. While the higher aspect ratio ring produces a more massive planet, it permits a more massive and therefore a stronger vortex, thus increasing its lifetime. However, since there are many effects at work here, from those determining the strength of the vortex when it separates from the planet to those which act to dissipate the vortex, the impacts of which are difficult to disentangle, this remains hypothetical and thus should be investigated further in future simulations.

Vortices can also form via the Rossby wave instability (RWI) at the edges of the gap created by the planet in the lower aspect ratio rings. This does not occur in the $h = 0.1$ rings or in the lowest mass $h = 0.05$ ring, as the planet cannot carve a deep enough gap to trigger this instability. These vortices are generally weak, and only the smaller dust grains, being strongly coupled to the gas, tend to form asymmetries; the larger dust grains, having longer stopping times, typically maintain their axisymmetric ring structures. Gap formation is even easier in the coldest rings, thus even the planet in the $50 M_\oplus$ ring is able to carve a gap whose edges are unstable to the RWI. However, only in the more massive rings are the vortices that form at the gap edges able to trap dust significantly. These vortices eventually dissipate, and the dust rings re-form within the axisymmetric pressure maxima, and thus have no impact on the planet or final disc structure.

\section{Discussion}
\subsection{Vortex-assisted pebble accretion}
\label{sec:vapa}
One of the key findings of \cite{cummins22} was that the jump in pebble accretion rate seen when the vortex initially forms around the planet can be interpreted as arising due to the vortex providing a larger accretion cross-section than the embryo's Hill sphere, with its maximum semiminor axis being $\approx 2H_\ug/3$. The resulting expression for the accretion rate,
\begin{equation}
  \dot{M}_\mathrm{peb, vort} \approx \frac{4}{3}\mathit{\Omega}_\mathrm{K} H_\ug^2 \mathit{\Sigma}_\mathrm{peb},
  \label{eqn:m_peb_dot_va}
\end{equation}
shows the dependence on the aspect ratio and dust surface density, which means that this hypothesis can be directly tested with the simulations we performed here. As the simulations with the same aspect ratio vary only by dust mass and therefore dust surface density in the ring, the accretion rates should vary linearly with surface density between simulations. While this is true for both the vortex-assisted pebble accretion rate and the standard pebble accretion rates, in the case of vortex-assisted accretion, the accretion rate is independent of planet mass, enabling a potential means of discriminating between the two accretion scenarios.
Similarly, simulations performed with the initial same dust mass vary only by the disc aspect ratio, though only the $150 M_\oplus$ rings provided an enhancement in pebble accretion rate due to the presence of the vortex, and only for aspect ratios $h(a) = 0.035$ and 0.05, thus this scaling is more difficult to reliably test with this set of simulations.

\begin{figure}
  \centering
  \includegraphics{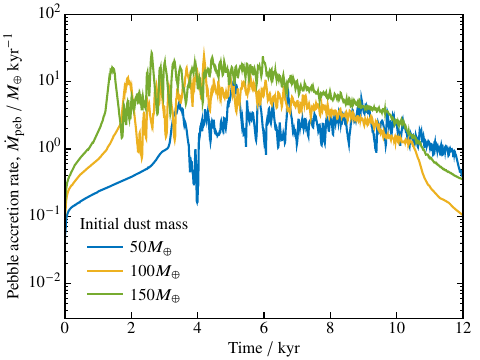}
  \caption[Pebble accretion rates for $h(a) = 0.035$.]{Pebble accretion rate onto the embryo growing in the $50 M_\oplus$, $100 M_\oplus$ and $150 M_\oplus$, $h(a) = 0.035$ rings over the first 12~kyr. Analytically, the pebble accretion rate is expected to scale linearly with the dust surface density and, therefore, the initial dust mass. The initial peaks in accretion rate due to vortex formation reach 5, 11 and $16 M_\oplus$~kyr$^{-1}$ respectively, which do scale approximately linearly with the initial dust mass, while the time-averaged accretion rates (2.6, 5.1 and $10.8 M_\oplus$~kyr$^{-1}$ respectively) during the vortex-assisted pebble accretion phase scale logarithmically with the initial dust mass.}
  \label{fig:par_0035}
\end{figure}

Figure~\ref{fig:par_0035} shows the total pebble accretion rates in the $50 M_\oplus$, $100 M_\oplus$, and $150 M_\oplus$ rings with $h = 0.035$ during the first 12~kyr of the simulations. These should scale linearly with the local dust surface density and thus increase with the initial mass of dust in the ring. Furthermore, according to equation~(\ref{eqn:m_peb_dot_va}), they should be constant with time during the vortex-assisted phase. The initial peak observed in each of the accretion rates due to vortex formation around the planet agrees well with this expectation, scaling accordingly with initial dust mass. In each case, the accretion rate drops significantly after this peak, before rising again on a timescale of $\sim 0.5$~kyr and reaching a roughly constant value. After a period of time, which is dependent on the initial dust mass, the pebble accretion rate begins to decay before sharply falling once the vortex separates from the planet. The steady decay in pebble accretion rate, particularly apparent in the $100 M_\oplus$ and $150 M_\oplus$ rings (starting around 6~kyr), will be discussed in the following sections. The time averages of the pebble accretion rate during the vortex-assisted phase do not scale linearly with the initial dust mass. These are $2.6 M_\oplus$~kyr$^{-1}$, $5.1 M_\oplus$~kyr$^{-1}$ and $10.8 M_\oplus$~kyr$^{-1}$ for the $50 M_\oplus$, $100 M_\oplus$ and $150 M_\oplus$ rings respectively. Thus, the pebble accretion rate in the most massive ring is higher than expected. Note, however, that the initial dust mass serves only as an estimate of the surface density of dust available for pebble accretion: higher pebble accretion rates will lead to stronger vortex formation, which in turn leads to higher surface density in the vicinity of the planet due to dust trapping by the vortex.

\begin{figure*}
  \centering
  \includegraphics{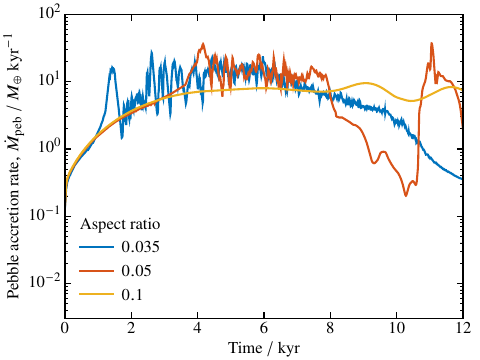}
  \includegraphics{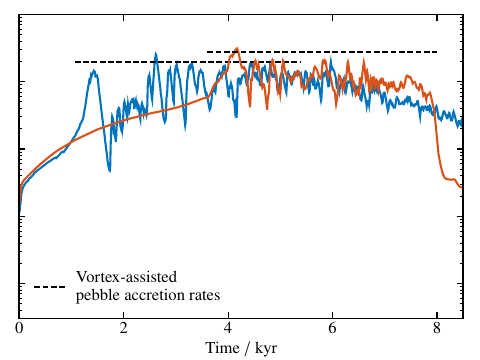}
  \caption[Pebble accretion rates in $150 M_\oplus$ rings.]{Pebble accretion rates in the $150 M_\oplus$ rings with aspect ratios of 0.035, 0.05 and 0.1 over the first 12~kyr of the simulations. Both the $h = 0.035$ and $h = 0.05$ rings lead to vortex formation, enhancing the accretion rate, while the thermal feedback is insufficient in the highest temperature ring. The right-hand panel shows the accretion rates of the centimetre-sized pebbles only, for the $h = 0.035$ and $h = 0.05$ rings over first 8~kyr, with the analytic time-averaged vortex-assisted pebble accretion rates.}
  \label{fig:par_150}
\end{figure*}

Figure~\ref{fig:par_150} shows the pebble accretion rates in the $150 M_\oplus$ rings over the first 12~kyr of the simulations. The accretion rates onto the embryo are very similar for the first 1000 years, as expected, before a vortex forms in the coldest disc and the accretion rate increases by a factor $\sim 10$ as a result. After 3.5~kyr, a vortex forms in the $h(a) = 0.05$ disc, and the accretion rate increases by a factor $\sim 5$. Given the difference in scale height, the accretion rates during the period while the vortex is co-located with the planet should vary by a factor $\approx 2$, which is not observed -- the accretion rates are approximately equal.
A direct comparison of these accretion rates is not straightforward, however. Vortex-assisted pebble accretion requires $2H_\ug^2/3 > R_\mathrm{H}^2$, which sets a limit on the planet mass for which this can occur at a given aspect ratio:
\begin{equation}
  M_\mathrm{P} \lesssim 3M_*\bigg(\frac{2h^2}{3}\bigg)^{3/2}.
  \label{eqn:m_va_hill}
\end{equation}
Beyond this mass, which is comparable to the thermal mass, standard pebble accretion in the Hill regime takes over as the planet's Hill radius exceeds the maximum semiminor axis of the vortex. For an aspect ratio of 0.035, this threshold is reached once the embryo grows to $\approx 47 M_\oplus$, which occurs shortly after 5~kyr in the $150 M_\oplus$ ring. In contrast, the vortex-assisted pebble accretion phase in the $h = 0.05$ ring lasts from 3.5 to 8~kyr, after which the vortex separates from the planet. The time-averaged pebble accretion rate during the vortex-assisted phase for the $h = 0.035$ ring, approximately $20 M_\oplus$~kyr$^{-1}$, is a factor of 1.4 lower than that during the equivalent phase in the $h = 0.05$ ring ($\approx 28 M_\oplus$~kyr$^{-1}$), rather than the predicted factor of 2. This can be seen in the right-hand panel of figure~\ref{fig:par_150}, which shows the accretion rate of the centimetre-sized pebbles onto the embryo and the time-averaged analytic accretion rates. Only the accretion rates for centimetre-sized dust grains are shown, as the Stokes number dependence is not accounted for in the analytic expression of the vortex-assisted pebble accretion rate, and since these largest grains have Stokes numbers closest to unity, this minimises the expected impact of this additional factor. However, while a discrepancy of order unity between the simulated and analytic accretion rates is not surprising, neglecting the Stokes number dependence should not impact the ratio between the accretion rates in different aspect ratio rings.

\begin{figure}
  \centering
  \includegraphics{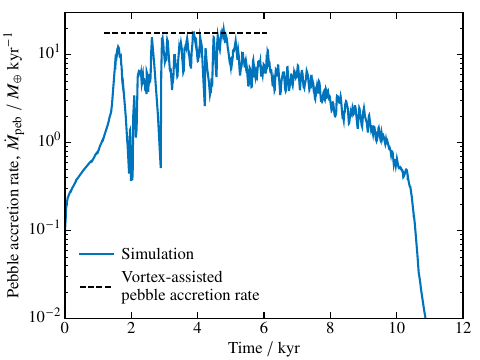}
  \caption[Pebble accretion rate in the $125 M_\oplus$, $h(a) = 0.035$ ring.]{Accretion rate of the centimetre-sized pebbles onto the embryo in the $125 M_\oplus$, $h(a) = 0.035$ during the vortex-forming phase. The blue line shows the pebble accretion rate from the simulation with thermal feedback. The black dashed line shows the analytic time-averaged vortex-assisted pebble accretion rate between the period when the vortex size exceeds the embryo's Hill radius.}
  \label{fig:par_125_0035}
\end{figure}

The fact that the embryo's growth in the more massive $h = 0.035$ rings transitions out of the regime where the vortex size exceeds the embryo's Hill sphere motivates taking a closer look at the pebble accretion rates during this period. Figure~\ref{fig:par_125_0035} shows the accretion rate of the centimetre-sized pebbles onto the embryo growing in the $125 M_\oplus$ $h = 0.035$ ring, for the first 12~kyr of the simulation, as well as the time-averaged pebble accretion rate given by equation~(\ref{eqn:m_peb_dot_va}) between the time that the vortex forms and the time at which the vortex no longer provides a greater accretion cross-section than the embryo's Hill sphere. Between $\sim 2$ and 5~kyr, the accretion rate shows large fluctuations but about a roughly constant value in time. After around 5~kyr the accretion rate begins to decline; coincidentally, this is around the time that the embryo's Hill radius exceeds the vortex size, but this is not a consistent outcome for all dust masses 
-- this decline begins earlier with increasing ring mass, except for in the most massive ring (see figures~\ref{fig:par_0035} and \ref{fig:par_150}). This can be explained by the fact that more massive rings generate a more massive planet. Still, the flat (time-averaged) accretion phase delays the point at which the decline in surface density offsets the increase in the planet's Hill radius.

\subsubsection{Vortex dynamics}
Given the large fluctuations observed in the pebble accretion rate while the vortex is co-located with the planetary embryo, it is worth considering how to interpret this ``vortex-assisted'' accretion rate. These fluctuations arise from the motion of the vortex relative to the embryo: their mutual interaction results in slight ($a\Delta \phi \lesssim 3$~AU) azimuthal oscillations in the vortex's position relative to the embryo. This moves dust in and out of the embryo's Hill sphere and thus causes fluctuations in the pebble accretion rate.

An additional component may be due to the growth and destruction of the vortex: as the vortex grows, it exerts stronger torques on the disc, exchanging angular momentum with it, which leads to radial migration. However, once it reaches its maximum size, the supersonic flow creates a shock, and the energy lost in this process weakens the vortex, which in turn slows down its migration. If the vortex remains in the vicinity of the embryo, the embryo continues to accrete, sourcing vortensity and thus allowing the vortex to grow again.

In this case, the vortex-assisted pebble accretion rate should estimate the accretion rate the vortex can provide when at its limiting size, and the fluctuations should average to this value. There will be a short duration over which the accretion rate reaches this value while the vortex is growing from the planet's Hill radius up to $2H_\ug/3$. Still, the accretion rate will likely exceed this rate as the vortex briefly exceeds this size, after which it loses energy and thus decreases in size, causing the pebble accretion rate to fall below this value. This process repeats until the vortex finally separates from the planet. This appears consistent with the accretion rates shown in, for example, figures~\ref{fig:par_150} and \ref{fig:par_125_0035}, when accounting for the missing Stokes number dependence in the analytic expression, which likely introduces a factor of $\sim 0.5$--0.7, given the Stokes numbers of the largest grains are $\sim 0.5$.

This then raises the issue of what drives the vortex to separate from the planet. One might expect the vortex to separate once the planet reaches its gap-opening mass due to the sudden change in the torques experienced by the vortex, causing it to migrate. In this case, the vortex should separate earlier for higher ring masses, as the gap-opening mass should be reached faster. However, there does not appear to be a clear trend with ring mass, as we now describe, considering the $h = 0.035$ rings.

In the $50 M_\oplus$ ring, the first vortex separates from the planet after 76 orbits, while the planet's mass is approximately $22 M_\oplus$, and the second shortly after (82 orbits). This planet's mass is comparable to the maximum mass reached in the simulations without thermal feedback (i.e. related to the isolation mass); the vortices still enter and cross the planet's Hill sphere until it has reached $\approx 26 M_\oplus$, after which they become confined to horseshoe orbits, indicating that gap formation is underway.
If we consider the thermal mass to be the mass at which the embryo's Bondi radius, $R_\mathrm{B} = GM_\mathrm{P}/c_\mathrm{s}^2$, equals the gas scale height, we find a thermal (and therefore gap-opening) mass of $\approx 28 M_\oplus$ -- this is close to the mass at which the vortices are confined to horseshoe orbits.

In the $75 M_\oplus$ ring, a small vortex separates at 40 orbits when the planet has reached $21 M_\oplus$, and the main vortex separates at 110 orbits, by which point the planet has reached $38 M_\oplus$. The first vortex is already confined to horseshoe orbits before the second vortex separates, indicating that the vortex remains co-located with the planet even after the planet has reached the gap-opening mass. Note also that this is approximately the mass at which the embryo's Hill radius exceeds the vortex size, thus the second vortex is contained within the embryo's Hill sphere and is therefore gravitationally bound to it. During this phase, the strength of the vortex decays until it ultimately separates from the planet.

In the $100 M_\oplus$ ring, a vortex separates from the embryo at 29 orbits ($M_\mathrm{P} \approx 20 M_\oplus$), while pebble accretion and thus vortex formation at the planet location persist. This vortex re-enters the embryo's Hill sphere upon their next encounter at 58 orbits and merges with the newly-formed vortex. The combined vortex separates from the planet at 69 orbits, once the planet has reached $47 M_\oplus$. Note that this is approximately equal to the mass at which Hill pebble accretion takes over from vortex-assisted, according to equation~(\ref{eqn:m_va_hill}).
In the $125 M_\oplus$ and $150 M_\oplus$ rings, the vortex separates after 101 orbits ($M_\mathrm{P} = 73 M_\oplus$) and 96 orbits ($M_\mathrm{P} = 87 M_\oplus$) respectively. In both of these cases, the planet's Hill radius exceeds the vortex size, while the vortex is still co-located with the planet. Thus, the vortex is contained within the planet's Hill sphere. The vortex therefore remains bound to the planet beyond the planet's gap-opening mass, however, the pebble accretion rates may be sufficiently high that the Hill sphere encloses the vortex faster than the vortex can separate. Clearly, gap-opening by the planet plays a role in the vortex's separation.  In future work, simulations with finer dust mass spacing may elucidate this.

An interesting feature of the pebble accretion rates is that, in the rings with properties most conducive to strong thermal feedback, before the vortex-assisted pebble accretion phase, they are slightly reduced with respect to the corresponding simulations ignoring the thermal feedback. Without any accretion luminosity, the embryo's location is a local density maximum (see, e.g. figure~3 of \citealt{cummins22}). However, as its accretion luminosity forms a hotspot, for the gas in its vicinity to maintain hydrostatic equilibrium, the increase in temperature due to the hotspot is accompanied by a reduction in gas density. As vortensity generation depends on the gas surface density and pressure gradients, this local pressure structure is critical -- the result is that there are two vortices that form, one ahead of and one behind the planet. Being pressure maxima, these slightly reduce the dust surface density in the embryo's Hill sphere until they merge to form one large vortex surrounding the planet.

\begin{figure*}
  \centering
  \includegraphics{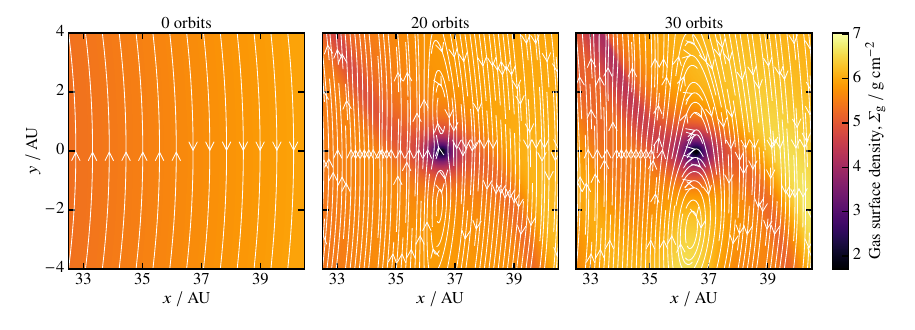}
  \caption[Vortex formation and merging]{Gas surface density and velocity streamlines showing the evolution of the vortex structure around the hotspot from the accreting planet, which is located at (36.5, 0). The snapshots are viewed in the frame rotating with the planet. Two vortices initially form, one ahead of and one behind the planet, which grow and merge to form one large vortex surrounding the planet.}
  \label{fig:vortex_formation}
\end{figure*}

To demonstrate this process, we performed a simulation in which the planet's mass was set to and fixed at zero, and artificially introduced a hotspot around the planet so that any feedback on the disc was purely due to the thermal structure and not the planet's gravity. Figure~\ref{fig:vortex_formation} shows three snapshots of the gas surface density from this simulation, overlaid with the gas velocity streamlines. This shows the vortex formation process, with vortices initially forming ahead of and behind the planet, which ultimately merge to form one large vortex.
This explains the edge-case simulations (e.g. $h(a) = 0.05$, $100 M_\oplus$ and $125 M_\oplus$), in which vortices were formed but unable to produce the distinctive enhancement in pebble accretion rate, as the vortices were not large enough to merge and form a single vortex surrounding the planet. The embryo's luminosity radius exceeded its Hill radius in each of the $h = 0.05$ discs, but only in the most massive ring did the vortices merge. While we have argued that $R_\mathrm{L} > R_\mathrm{H}$ for vortex formation, figure~\ref{fig:rlrh_fixedh} suggests that $R_\mathrm{L} \gtrsim 3 R_\mathrm{H}$ to trigger vortex-assisted pebble accretion. Closer inspection of the vortensity structure around the embryo as the luminosity radius grows suggests that the luminosity radius must extend to the centres of each of the vortices for them to merge.

The generation of vortensity in these two regions may also contribute to the oscillations of the vortex about the planet's location: since vortensity is generated at either end of the single, merged vortex, as the smaller vortex structures become incorporated into the larger vortex its morphology will likely vary. Thus, the small-scale vortex dynamics near the accreting embryo are complex, and future simulation work can be used to study this process in detail. 

\subsection{Pebble isolation masses in axisymmetric dust traps}
\label{sec:m_iso}
To understand what sets the final masses for planetary embryos forming in pressure-generated dust rings, it is useful first to consider the final masses reached in the simulations, ignoring the embryo's accretion luminosity, to isolate the effects of the thermal feedback on the disc.
Based on the pebble isolation mass concept, one should expect the pebble accretion rate (without thermal feedback) to drop off once the planet reaches a critical mass, perturbing the disc and thus preventing further accretion. Formulations of the pebble isolation mass show a scaling with the cube of the aspect ratio, and the results obtained here do indeed show that larger planets form in higher aspect ratio discs for all initial dust masses. However, there is also a dependence on the dust mass in the ring.

\cite{bitsch18} determined empirical scaling laws for the pebble isolation mass as functions of viscosity and the pressure gradient, finding
\begin{equation}
  \begin{split}
    M_\mathrm{iso} = 50 M_\oplus & \bigg(\frac{h}{0.05}\bigg)^3 \bigg[0.34 \bigg(\frac{-3}{\log_{10}\alpha}\bigg)^4 + 0.66\bigg] \\
    & \times \bigg[1-\frac{1}{6}\bigg(\pdbyd{\log P}{\log r} + 2.5\bigg)\bigg].
  \end{split}
  \label{eqn:m_iso_B18_2}
\end{equation}
Note that we have scaled the pre-factor with the stellar mass used here (using the physical approach in \citealt{bitsch18}), as the pebble isolation mass is expected to scale linearly with stellar mass \cite[e.g.][]{liu19a}. For a ring with an aspect ratio of 0.035 and zero pressure gradient\footnote{Although the simulations performed by \cite{bitsch18} were 3D, they considered a gas surface density profile $\mathit{\Sigma}_\ug \propto r^{1/2}$, which provides a local maximum in vertically-integrated pressure if $h \propto r^{1/4}$. Note also that they found differences in pebble isolation mass of up to a factor of 1.5 lower in 2D than in 3D simulations.} (by the nature of it being a dust trap), the pebble isolation mass according to this formulation is $M_\mathrm{iso} \approx 7 M_\oplus$. In comparison, the thermal mass is $M_\mathrm{th} \approx 85 M_\oplus$. Neglecting thermal feedback, the embryo reaches just over $14 M_\oplus$ in the least massive ring with this scale height and approximately $23 M_\oplus$ in the most massive ring.

The pebble isolation masses, given by equation~(\ref{eqn:m_iso_B18_2}), for aspect ratios of 0.05 and 0.1 are $22 M_\oplus$ and $179 M_\oplus$ respectively. In the $h(a) = 0.05$ rings, the final planet masses all exceed this isolation mass, while in the $h(a) = 0.1$ rings, the planet cannot reach this mass as it exceeds the total dust mass in the simulation. While figure~\ref{fig:paramstudy_summary} shows that more massive rings enable more massive planets to form, the pebble accretion rates (e.g. figure~\ref{fig:mp_h_nh_fixedh}) show that accretion ceases earlier in more massive rings, as planets reach their isolation mass earlier. The subsequent increase in planet mass is due to the finite time it takes to open the gap.

The situation is further complicated by the fact that even though the planet doesn't reach the pebble isolation mass predicted by equation~(\ref{eqn:m_iso_B18_2}) in the $h = 0.1$ rings, it does cease accretion. While there is a shallow dependence on ring mass for the smallest aspect ratios, it is almost linear for $h(a) = 0.1$. The dust mass fraction accreted from the ring decreases with increasing dust mass, from 95\% in the least massive ring to 71\% in the most massive ring. This suggests that the accumulated torque on the disc from the planet throughout its growth is important, rather than there being a well-defined critical mass below which the planet can accrete, and beyond which no accretion can take place.

In general, larger scale heights result in a higher final planet mass because the mass required for gap-opening increases with the disc aspect ratio. In the simple picture of pebble accretion without any thermal feedback on the disc, the planet will simply accrete until it begins to open a gap, and pebbles can no longer drift into its orbit, becoming trapped at the pressure maximum at the outer gap edge. While this expectation is verified in our simulations, there is an additional consideration in that the planet is accreting in a dust trap, the width of which is set by drift-diffusion equilibrium. This width is independent of the disc aspect ratio. However, the width of the gap opened by the planet depends strongly on the aspect ratio, as the planetary wakes form shocks which deposit angular momentum further from the planet with increasing aspect ratio \citep{goodman_rafikov01}. Each dust species becomes more strongly trapped as their Stokes numbers increase towards unity. With the gas pressure profile used in these simulations, the millimetre grains are trapped in a ring of FWHM $\approx 7$~AU, the 3~mm grains within 4~AU and the centimetre grains within 3~AU. Furthermore, the fact that the dust grains have an MRN size distribution means that most of the mass is in large grains, which are the most strongly trapped and have the highest pebble accretion rate.

Table~\ref{tab:gap_widths} lists the widths of the gaps formed by the planet in each simulation after 60~kyr, i.e. after the planets have all reached their final mass. The gap widths are reported as full-width at half maxima, where in each case, the maximum is the difference between the top and bottom of the gap and the full width is measured at the midpoint of these values. The ``top'' refers to the average of the local maxima on either side of the gap, and the ``bottom'' refers to the average of the local minima immediately on either side of the planet. The planets in the rings with an aspect ratio of 0.1 do not form gaps within the simulated time. The planet in the most massive ring may eventually form a deep gap, but since it has ceased accreting by this point, gap formation would not impact the results presented here. In the rings with aspect ratios of 0.035 and 0.05, each planet forms a gap: the least massive planet, i.e. that which formed in the $50 M_\oplus$ ring, forms a gap that is comparable in width to the extent of the millimetre grains, but in all other rings the planet forms a gap that is wider than the dust ring. The planets can therefore grow beyond the classical pebble isolation mass as the mass within the dust trap already exceeds this, and the planet can accrete a significant fraction of this mass before it begins to repel material onto horseshoe orbits and dust can no longer enter its Hill sphere to be accreted.

\begin{table*}
	\centering
	\caption[Gap widths]{Gap full-width half-maxima after 60~kyr. All widths are in units of AU. Values in parentheses are for simulations without thermal feedback. Gap widths for the $h(a) = 0.1$ rings were not possible to measure due to their negligible or shallow depths.}
	\label{tab:gap_widths}
  \begin{tabular}{@{}ll|ccccc@{}}
    &       & \multicolumn{5}{c}{Ring mass $\:/\:M_\oplus$}\\
    &       &     50    &     75     &    100    &    125     & 150\\
		\hline
    \multirow{3}{*}{\rotatebox{90}{$h(a)$}}
    & 0.035 & 7.8 (6.8) &  8.5 (7.2) & 9.4 (7.4) &  9.9 (7.5) & 10.5  (7.6) \\
    & 0.05  & 9.4 (9.4) & 10.0 (9.6) & 9.7 (9.7) & 10.3 (9.9) & 12.2 (10.1) \\
    & 0.1   & -- (--)   & -- (--)    & -- (--)   & -- (--)    & $\sim 15$ ($\sim 15$) \\
  \end{tabular}
\end{table*}

As each of the planets in the $h = 0.035$ rings generated strong thermal feedback on the disc, these provide a means of understanding what role this plays in setting the final mass. As discussed above, the planet continues to accrete past the mass attained without thermal feedback due to the presence of the vortex around the planet. The masses at which the accretion rate rapidly drops off (e.g. around 9~kyr in figure~\ref{fig:par_100_0035}) also vary with the initial dust mass, as these correspond to the times when the vortex separates from the planet. For the most massive ring, this occurs when the planet has reached $83 M_\oplus$, and for the least massive ring, this occurs when the planet has reached $22 M_\oplus$. These significantly exceed the mass reached without thermal feedback, where the accretion rate begins to fall off once the planet reaches the thermal mass.

The gap widths are increased when thermal feedback is included, though only because the planet is more massive. Torques from the planet dominate gap-opening rather than torques from vortices.
As evidenced by the $50 M_\oplus$ and $75 M_\oplus$ simulations, reaching the thermal mass appears to be important in the vortex separating from the planet. However, if the dust mass in the ring is high enough that the embryo can continue to accrete and source vortensity, and the aspect ratio is low enough that the embryo's Hill radius can exceed the vortex size, the vortex remains bound to the planet until the gap is opened and the vortex ultimately separates. The growth timescale and gap-opening timescale are important, and the growth history of the planet also plays a role: the $150 M_\oplus$ ring produces a more massive planet than the $125 M_\oplus$ ring, and the vortex separates from the planet at a higher mass.

Thermal feedback also significantly affects the planet's growth in the $h = 0.05$, $150 M_\oplus$ ring. The higher aspect ratio ring generates a more massive planet than in the $h = 0.035$ ring of the same mass, corroborating that larger scale heights result in higher masses, even when vortex assistance is involved, as the vortex is larger and can thus accumulate dust from a larger distance. Furthermore, at larger scale heights, gap-opening requires a higher planet mass, a longer gap-opening timescale, and a wider gap formed.

\subsection{Consequences for planet formation in dust rings}
One of the major challenges in planet formation is how to grow the cores onto which gas accretion can occur \citep[e.g.][]{Armitage2018}. The discovery that protoplanetary discs are not smooth but often structured with dust highly concentrated in local substructures \citep[e.g.][]{Andrews2020}, containing masses $\gtrsim 10 M_\oplus$ of solids, has yielded a potential solution to the first problem: how to bring sufficient solids together to form a planetary core. Given the majority of these substructures take the form of narrow axisymmetric rings \citep[e.g.][]{VanderMarel2019}, likely created by pressure trapping \citep[e.g.][]{dsharp6,rosotti20}, understanding the formation and evolution of planetary embryos in these environments is critical to our understanding of planet formation. Given the raw efficiency of pebble accretion, the fact that we actually see massive rings implies that pebble accretion does not continue unabated \citep{owen17,Lee2022,Lee2024}; otherwise, all the dust trapped would be rapidly accreted onto the embryo. Thus, there must be feedback that either prevents planet formation from occurring to start with, or that truncates planet formation before the dust in the ring is exhausted. 

\citet{Lee2022} and \citet{Carrera2022} argued that planetary seed formation may be more difficult than originally envisaged in dust traps, with a dependence on particle size and pressure bump properties. \citet{Lee2022} showed that clumping is easier at smaller separations, where the disc's scale height is lower. As we have shown, thermal feedback is more important at smaller scale heights, permitting larger planet masses, where the accretion enhancement is considerably larger. Thus, the planets that form in our simulations could be the seeds from which gas giants form. Since many close-in giant planets, which are believed to form further out in their parent discs \citep[e.g.][]{Dawson2018}, are extremely metal-rich \citep[e.g.][]{thorngren16}, then solid accretion, enhanced by thermal feedback, could play a role in explaining their origins. The advent of exoplanet characterisation with the James Webb Space Telescope will allow the atmospheres of close-in giant planets to be studied in detail, potentially elucidating the origins of their solid accretion \citep[e.g.][]{Kirk2024,Penzlin2024}.

\subsection{Caveats}
Given our goal was to isolate the impact of thermal feedback, we performed a small set of simplified simulations. Thus, it is worth discussing those simplifications and their possible impact. Firstly, we have ignored the back-reaction of the dust on the gas; as discussed in \citet{cummins22}, our most massive dust ring does not exceed a $\mathit{\Sigma}_\ud/\mathit{\Sigma}_\ug$ of unity; however, the 150~$M_\oplus$ has a maximum value of $\approx 0.8$ in our initial condition. Therefore, the back-reaction will not be negligible; however, given we see the importance of thermal feedback for the lowest surface densities down to a dust ring mass of 50~$M_\oplus$, the back-reaction will not change the basis of our results. \textcolor{mod}{Perhaps the largest uncertainty with the back-reaction is its impact on vortex evolution since large amounts of dust can become trapped in the vortex, and can exceed a dust-to-gas ratio of unity in the vortex core \citep[e.g.][]{lyra_lin13}}. Using 2D simulations there has been the suggestion that the vortex lifetime becomes reduced \citep[e.g.][]{fu14,raettig15}. However, this result is believed to be an artefact of 2D simulations, with 3D simulations showing that their lifetimes are unaffected by dust accumulation \citep[e.g.][]{lyra18, raettig21}. Thus, now the impact of thermal feedback has been isolated, the role of dust feedback should be considered in the future. 

Our current simulations ignored planetary migration. Simulations by \citet{guilera_sandor17} and calculations by \cite{morbidelli20} have shown that migration inside a dust trap tends to result in minimal migration, and the planet remains close to the peak of the dust distribution, continuing to accrete rapidly, thus resulting in similar final planet masses. However, \citet{Pierens2024} performed simulations including both thermal feedback and migration, although they parametrized the solid accretion. This work showed that while the planet could accrete at a sufficiently high rate to trigger thermal feedback and vortex formation, it did excite the eccentricity of the planet, which could potentially allow the planet to escape from the dust ring. Similar results were found by \citet{Chrenko2023} who found accreting luminous low-mass planets could escape from migratory traps at pressure bumps, while \citet{Velasco2024} showed that some of the eccentric orbits induced allowed the planet to continue to accrete from the dust trap. \citet{Pierens2024} also found dust feedback to be reduced in a weaker interaction between the formed vortices and the forming planet, allowing the planet to remain in the dust ring, even with an eccentric orbit. 

Thus, while our general results are robust to our simplifications, future simulations should include both migration and the dust's feedback on the gas while resolving pebble accretion within the simulation. Furthermore, while we've attempted to mimic the impact of dust growth and fragmentation in the initialisation of our dust size distribution, it is becoming clear that the dust's size has a large impact on planetary accretion in dust traps \citep[e.g.][]{Lee2022,Pierens2024}; therefore, future work should consider whether dust growth and fragmentation can occur on sufficiently fast timescales to reset the size distribution during the planet's rapid growth. If this is possible, it could continue to supply the larger grains, which are most efficiently accreted, thereby enhancing thermal feedback further.    

\section{Conclusions}
In this study, we have investigated the impact of a planetary embryo's accretion luminosity on its growth via pebble accretion in an axisymmetric dust trap for a range of dust masses and disc temperatures. We have shown that the process of vortex formation via the thermal feedback on the disc from the embryo's accretion luminosity is robust and can significantly affect a planet's growth.
Under the most favourable conditions for strong thermal feedback, the planet's mass can be increased by a factor of four with respect to the same initial conditions but neglecting the thermal feedback. This effect is most pronounced in dust traps located in cold regions of discs, such that their aspect ratios are $\lesssim 0.05$; otherwise, large amounts of dust -- in excess of $100 M_\oplus$ -- are required in the trap to have an impact. 

The rings considered here as the formation sites of these planets were motivated by those observed in the DSHARP survey, with aspect ratios between 0.035 and 0.1 and dust masses between $50 M_\oplus$ and $150 M_\oplus$. In the discs with the highest aspect ratio, the background temperature is relatively high, and the thermal feedback was therefore not strong enough to source vortensity sufficient for vortex formation, and ultimately had no impact on the embryo's growth. In the 0.05 aspect ratio rings, only those with the highest mass provided an accretion rate sufficient to produce strong thermal feedback. Of the rings studied by \cite{dsharp6}, only one had an estimated scale height as low as 0.05, which was also the least massive, thus this process is not likely to take place within those rings. Given the short timescales on which the pebble accretion and vortex formation processes take place, one could plausibly speculate that this process has already taken place in some of the observed ringed discs and that the resulting planet is as yet undetected, residing between the rings formed as a consequence. Of course, an unambiguous observational signature would be required to make this claim. The final state of the disc in our simulations is typically two rings and dust trapped in L$_5$, which is due only to the process of planet formation rather than vortex formation during the pebble accretion phase. As discussed in \citet{cummins22}, the fact gas drag modifies the preferential trapping location in L$_5$ in a particle size dependant way, it provides an observational route to distinguish between L$_5$ trapping and vortices as the origin for observed asymmetries. 

However, the vortex formation phase impacts the size distribution of dust, with a significant depletion of centimetre-sized ($\mathrm{St} \sim 0.5$) dust grains. Multi-wavelength observations of discs can constrain the sizes of dust grains present, although these are still subject to uncertainty regarding the physical properties of the dust grains. Similarly, both the maximum possible grain size (e.g. the fragmentation size) and their Stokes numbers are difficult to determine without knowledge of gas densities and turbulence levels.

In the discs with the lowest scale height considered, even the least massive ring provided a pebble accretion rate which led to vortex formation. This motivates further investigation into the ring mass at which the pebble accretion rate is insufficient to lead to vortex formation.

We have also shown that planet formation in a pressure-enabled dust trap can lead to planets with masses in excess of the classical pebble isolation mass, if the dust is confined to a ring more narrow than the gap opened by the planet. This is especially so for high pebble accretion rates, such that the growth timescale is shorter than the gap-opening timescale.
If the radial extent of the dust within the ring is less than the gas pressure scale height, the entire dust distribution will be contained within the gap opened by the planet, since the spiral wake drives shocks at a distance $\approx H_\ug$, creating a gap of order a few times the scale height. A long-lived pressure bump cannot be narrower in radial extent than $\sim H_\ug$, otherwise it cannot establish vertical hydrostatic equilibrium and can also become Rossby-unstable \citep{ono16, dsharp6}. However, the radial extent of pebbles within a pressure bump is likely to be significantly narrower than the gas, given their expected Stokes numbers. The rings analysed in the DSHARP survey were estimated to have dust widths $w_\ud \sim 0.5 H_\ug$ to $2 H_\ug$. Thus, this effect could be important for consideration in planet formation models, which may otherwise assume that pebble accretion stops at the classical isolation mass.

Regarding the prospect of the ring re-forming once the vortex dissipates, our results suggest that even in the least massive rings, the planets produced trap dust released by the vortex in L$_5$. \textcolor{mod}{The aspect ratio of the vortex formed is $\sim 3.5$, making it susceptible to elliptical instabilities \citep[e.g.][]{Lesur2009}. This instability could shorten the vortex's lifetime and thus lead to earlier release of dust, if the vortex is destroyed before the planet establishes its co-orbital region. However, the timescale on which the vortex decays through interactions with the planet is more significant, as this effect in itself weakens the vortex and thus the strength of any elliptical instabilities. Higher viscosities may also mitigate dust trapping in L$_5$, as a more massive planet would be required. However, higher viscosities may also prevent vortex formation in the first place, as they would widen the dust ring, reducing the dust surface density and, therefore, the pebble accretion rate. Similarly, a higher viscosity has consequences for the initiation of the streaming instability to form the embryo: estimating the pebble scale height as $H_\ud / H_\ug \sim \sqrt{\alpha/\mathrm{St}}$, the vertical extent of pebbles is greater for more turbulent discs, which may prevent the dust-to-gas ratio reaching the required value of unity.}
This therefore reinforces the need to include planetary migration in future simulations, as the interaction between the embryo and vortex could cause the planet to migrate away from the vortex \citep{Pierens2024}, therefore allowing the ring to re-form once the vortex dissipates.

By examining the disc properties in the planet's vicinity during the vortex formation phase, we have shown that the pressure structure generates a baroclinic term that forms two vortices -- one ahead of and one behind the planet. Only if the accretion luminosity of the planet is sufficiently high for these two vortices to merge will the vortex-assisted pebble accretion phase proceed. We have hypothesised that the luminosity radius must be large enough for the vortices to merge -- specifically that the luminosity radius must extend to the centres of these two vortices rather than simply being larger than the embryo's Hill radius. 

Finally, we have attempted to determine the factors contributing to the dynamics of the vortex during its formation and while it is co-located with the planet, as well as what causes it to ultimately separate from the planet. The embryo reaching the thermal mass can cause vortex migration; however, around this mass, the embryo's Hill radius exceeds the vortex's radial extent, causing the vortex to become bound to the planet for an extended period. What ultimately drives the vortex to separate after this phase remains unclear, but it may be possible to determine by performing simulations at a fixed aspect ratio with more closely spaced ring masses.

\section*{Acknowledgements}
We are grateful to the anonymous referee for comments which improved the manuscript. DPC was supported by a 2017 and 2020 Royal Society Enhancement Award. JEO is supported by a Royal Society University Research Fellowship. This project has received funding from the European Research Council (ERC) under the European Union’s Horizon 2020 research and innovation programme (Grant agreement No. 853022, PEVAP).
This work was performed using the DiRAC Data Intensive service at Leicester, operated by the University of Leicester IT Services, which forms part of the STFC DiRAC HPC Facility (www.dirac.ac.uk). The equipment was funded by BEIS capital funding via STFC capital grants ST/K000373/1 and ST/R002363/1 and STFC DiRAC Operations grant ST/R001014/1. DiRAC is part of the National e-Infrastructure.
This work was performed using the Cambridge Service for Data Driven Discovery (CSD3), part of which is operated by the University of Cambridge Research Computing on behalf of the STFC DiRAC HPC Facility (www.dirac.ac.uk). The DiRAC component of CSD3 was funded by BEIS capital funding via STFC capital grants ST/P002307/1 and ST/R002452/1 and STFC operations grant ST/R00689X/1. DiRAC is part of the National e-Infrastructure.
For the purpose of open access, the authors have applied a Creative Commons Attribution (CC-BY) licence to any Author Accepted Manuscript version arising.

\section*{Data availability}
The data underlying this article will be shared on reasonable request to the corresponding author.




\bibliographystyle{mnras}
\bibliography{references} 




\appendix




\bsp	
\label{lastpage}
\end{document}